\documentclass[conference]{IEEEtran}
\ifCLASSINFOpdf
\else
\fi

\usepackage{amsmath,amsfonts}

\usepackage{bbding} %对勾叉号
\usepackage{pifont}

\usepackage{makecell}
\usepackage{colortbl} % table color

\usepackage{tikz}
\usetikzlibrary{patterns}

\usetikzlibrary{arrows.meta}
\usetikzlibrary{automata,positioning}

\usepackage{threeparttable}

\renewcommand*{\arraystretch}{1.5}%
\definecolor{tabred}{RGB}{230,36,0}%
\definecolor{tabgreen}{RGB}{0,116,21}%
\definecolor{taborange}{RGB}{250,124,30}%
\definecolor{tabbrown}{RGB}{171,70,0}%
\definecolor{tabyellow}{RGB}{251,253,169}%
\newcommand*{\vcorr}{%
  \vadjust{\vspace{-\dp\csname @arstrutbox\endcsname}}%
  \global\let\vcorr\relax
}% 

\usepackage{mathtools}

\usepackage{lscape}
\usepackage[figuresright]{rotating}
\usepackage[graphicx]{realboxes}
\usepackage{adjustbox}
\usepackage{caption}

\usepackage{subfigure}

\usepackage{colortbl} 
\usepackage{xfrac}

\usepackage{float}

\usepackage{fbox} %use box
\usepackage{fancybox}%use other types of boxes
\usepackage{framed}

\usepackage{multirow}

\usepackage{CJKutf8}  
\usepackage{pgfplots}
\usepackage{filecontents}

\usepackage{hyperref}

\usepackage{tabularx,makecell, caption}
\usepackage{enumitem} % to control Itemization spacing

\newcolumntype{L}{>{\arraybackslash}X}

\usepackage{multicol}
\usepackage{listings}
\usepackage{blindtext}
\usepackage{etoolbox,xstring,mfirstuc,textcase}
\usepackage{listings}

\usepackage{subfiles}
\usepackage[most]{tcolorbox}

\usepackage{xcolor}
\newtheorem{defi}{Definition}

\usepackage{siunitx}
\usepackage{comment}
\usepackage{indentfirst} 
\usepackage{framed} 

\usepackage[font=small,labelfont=bf,tableposition=top]{caption}
\usepackage[normalem]{ulem}
\usepackage{booktabs}
\usepackage{dashbox}
\usepackage{xcolor}

\usepackage{listings}
\newenvironment{packeditemize}{
	\begin{list}{$\bullet$}{
			\setlength{\labelwidth}{4pt}
			\setlength{\itemsep}{0pt}
			\setlength{\leftmargin}{\labelwidth}
			\addtolength{\leftmargin}{\labelsep}
			\setlength{\parindent}{0pt}
			\setlength{\listparindent}{\parindent}
			\setlength{\parsep}{0pt}
			\setlength{\topsep}{1pt}}}{\end{list}}

\newenvironment{takeaway}{
\MakeFramed{\advance\hsize-\width\FrameRestore}}
{\endMakeFramed}

\newenvironment{takeaway1}{

\MakeFramed{\advance\hsize-\width\FrameRestore}}
{\endMakeFramed}

\usepackage{framed}
\definecolor{formalshade}{rgb}{0.95,0.95,0.97}
\definecolor{darkblue}{rgb}{0.14,0.22,0.52}

\usepackage{url}

\usepackage{color}

\usepackage[linesnumbered,ruled,vlined]{algorithm2e}
\usepackage{algpseudocode}

\begin{document}

\title{Slow is Fast! Dissecting Ethereum's \\ Slow Liquidity Drain Scams}

\newcommand{\acro}{{SLID}~}
\newcommand{\nasrin}[1]{\textcolor{red}{#1}}
\newcommand{\minh}[1]{\textcolor{blue}{#1}}
\newcommand{\minhq}[1]{\textcolor{orange}{#1}}
\newcommand{\qw}[1]{\textcolor{magenta}{#1}}
\newcommand{\jas}[1]{\textcolor{blue}{{#1}}}
\newcommand{\bk}[1]{\textcolor{purple}{{#1}}}

% \author{Anonymous Author(s)}

 \author{
  {\rm Minh Trung Tran$^\dag$, Nasrin Sohrabi$^\ddag$, Zahir Tari$^\dag$, Qin Wang$^\S$, Minhui Xue$^\S$ , Xiaoyu Xia$^\dag$}\\
  $^\dag$RMIT University $|$ $^\ddag$Deakin University $|$ $^\S$CSIRO Data61, Australia
%  \and
 % {\rm Nasrin Sohrabi}\\ 
%  Deakin University
 % \and
 % {\rm Zahir Tari}\\
 % RMIT University
 % \and
 % {\rm Qin Wang}\\
 % {\rm CSIRO}
 % \and
 % {\rm Minhui Xue} \\
 % {\rm CSIRO}

  }
\IEEEoverridecommandlockouts
\makeatletter\def\@IEEEpubidpullup{6.5\baselineskip}\makeatother
\IEEEpubid{\parbox{\columnwidth}{
		Network and Distributed System Security (NDSS) Symposium 2026\\
		23-27 February 2026, San Diego, CA, USA\\
		ISBN 979-8-9894372-8-3\\
		https://dx.doi.org/10.14722/ndss.2025.[23$|$24]xxxx\\
		www.ndss-symposium.org
}
\hspace{\columnsep}\makebox[\columnwidth]{}}

\maketitle

\begin{abstract}
We identify the \textit{slow liquidity drain} (SLID) scam, an insidious and highly profitable threat to decentralized finance (DeFi), posing a large-scale, persistent, and growing risk to the ecosystem. Unlike traditional scams such as rug pulls or honeypots (USENIX Sec'19, USENIX Sec'23), \acro gradually siphons funds from liquidity pools over extended periods, making detection significantly more challenging. 
In this paper, we conduct the first systematic measurement study of SLID scams among 319,166 liquidity pools across six major decentralized exchanges (DEXs). We identified 3,117 \acro affected liquidity pools, resulting in cumulative losses of more than US\$103 million. We propose a rule-based heuristic and an enhanced machine learning model for early detection. Our machine learning model achieves a detection speed 4.77 times faster than the heuristic while maintaining the F1-Score of 0.93. Our study establishes a foundation for protecting DeFi investors at an early stage and promoting transparency in the DeFi ecosystem.
\end{abstract}

%\begin{IEEEkeywords}
%Ethereum, Liquidity drain, Scam detection, Rug pull, Honeypots
%\end{IEEEkeywords}

% For peer review papers, you can put extra information on the cover
% page as needed:
% \ifCLASSOPTIONpeerreview
% \begin{center} \bfseries EDICS Category: 3-BBND \end{center}
% \fi
%
% For peerreview papers, this IEEEtran command inserts a page break and
% creates the second title. It will be ignored for other modes.
\IEEEpeerreviewmaketitle

\section{Introduction}~\label{sec:introduction}

Decentralized Finance (DeFi) \cite{sokdefi,jiang2023decentralized,defiintro} has emerged as an innovative landscape with the potential to revolutionize the global financial system~\cite{defia1,defia2}.  DeFi applications aim to replace traditional financial services by eliminating intermediaries and promoting transparency and permissionlessness through blockchain technology \cite{blockchainad}.

A pivotal moment for DeFi occurred with the rise of Uniswap \cite{uniswap1}, the leading DEX on Ethereum. Uniswap introduced the Automated Market Maker (AMM) \cite{dex1,dexamm,dexintroduction}, enabling users to create \textit{liquidity pool}s (LPs)\footnote{Within this paper, we use the abbreviation \textbf{LP} to denote the \textit{liquidity pool}, contrasting with sources \cite{sokdefi,cousaert2022sok} that define it as the \textit{liquidity provider}.} using their newly created tokens and facilitating direct asset exchanges without reliance on third parties like centralized exchanges (CEXs). Subsequent DEXs, such as Curve \cite{crv0} and SushiSwap~\cite{sushi0}, further attracted investors by offering a broad range of cryptocurrencies, including SHIB \cite{shib} and PEPE \cite{pepe}. The market value of DEXs \cite{unitrending,shibatrend,pepetrend} surged from \$587K in 2018 to \$7.71B during the “DeFi summer” of 2020 and eventually surpassed \$100B in 2021 \cite{defivalue}. To date, DEXs handle billions of dollars in daily trading volume \cite{defidailyvol}.

DEXs unsurprisingly became prime targets for financial scams \cite{dexscam1,dexscam2}. De.fi reported \cite{dedotfi2023} a staggering \$1.39b loss in Ethereum's DeFi sector due to scams.
These scams often involve the rapid creation of LPs, allowing fraudsters to retain full control over monetary value. Once invested, users are unable to sell or recover funds. Two infamous examples include \textit{rug pull} \cite{rugpull} (developers abruptly abandon a project after withdrawing liquidity), and \textit{honeypot} \cite{honeypot} (luring users by permitting deposits but blocking withdrawals).

Although increased user awareness reduced the prevalence of such straightforward scams, fraudsters are now resorting to more sophisticated schemes designed to bypass detection mechanisms and exploit unsuspecting users. Consider the case of a GoPlus report that flagged the Emily token as a potential exploitation due to the presence of an ``unlimited mint function''~\cite{goplusemily}. However, upon closer examination, the token did not display typical characteristics that are close to any of the existing DeFi scam types.

This observation demanded further scrutiny, leading to this paper discovering a far more pervasive threat which we term the \textit{slow liquidity drain} (SLID) scam. Unlike traditional DEX scams that exhibit clear characteristics (e.g., abrupt withdrawal, sale restriction), SLID operates without disrupting the normal functionality of LPs. Instead, it gradually siphons assets through a slow but calculated profit-taking strategy. This stealthy approach enables the scam to blend with legitimate LPs, making detection more challenging. Investors may experience prolonged losses without immediate realization.

Our analysis exposes the substantial fraudulent profits generated by \acro scams. Between 2018 and 2022, SLID scams siphoned off over \$64.33M in realized illicit gains, accounting for 14.3\% of the total DEX scam losses on Ethereum and Binance Chain combined (estimated at \$400M~\cite{sokdefiattack}). This striking figure underscores the broad impact of SLID scams within the DeFi ecosystem. Extending the scope to cover 2018 through 2024, these scams have inflicted a cumulative loss of \$103M, including realized and unrealized profits, marking them as a persistent threat to DeFi. Furthermore, our analysis of the dataset collected from Ethereum~\cite{eth} reveals a yearly increase in the number of \acro scams since their emergence in 2018 (Figure~\ref{tab:slid_lps}). This trend highlights the urgency of raising awareness about \acro scams to mitigate the growing threat they pose to DeFi ecosystems.

%\textcolor{gray}{Action 10: Our measurement approach provides concrete data-driven insights into the prevalence, financial impact, and distinctive operational patterns of SLID scams, thus filling a critical research gap in the existing literature and supporting future detection mechanisms}. 
% \jas{Can you provide any evidence with several citations, like anything on news or media?} 
% \nasrin{this is a novel scam and there is not official report about SLID.} 
%\jas{If we cannot observe something similar in the real world, it will lack endorsement for reviewers to support the argument as we cannot verify SLID. Evidence is mandatory!} \nasrin{there are some cases of SLID, which was not detected by any existing methods, but was reported by chainabuse, later our model detected it as slid. }

%jas{what is more insteresting is how you identidied SLID with reference to something similar reporeted in the real world, so this is the logic that you convince reviewers}
%\jas{I asked gpt to find me one. gpt showed me this one. Is that similar? https://www.wired.com/story/memecoin-kid-backlash/}

We progress by addressing the following questions.
%\textbf{This paper} introduces a novel and largely unexplored high-profit tactic in the DeFi scam on DEXs — the {\it slow liquidity drain} (SLID) scam. We present the first comprehensive study of this newly identified fraud and propose effective detection methods. Unlike traditional DEX scams such as rug pull (characterized by abrupt withdrawal or sell) or honeypot (blocking users from selling through malicious smart contracts), the SLID scam operates without disrupting the normal function of a LP. Instead, it gradually siphons off assets through a slow and methodical profit-taking strategy, allowing it to blend in with legitimate LPs while stealthily draining investor funds, making it difficult to detect.
%The research is conducted by addressing the following research questions:
%Can we identify any mutual patterns amongst the fraudulent instances for detection purposes?
%How can we build a more robust method to detect \acro scams at an early stage?
\begin{packeditemize}
    \item \textbf{RQ1.} \textit{What are the common features of \acro scams? } \acro scams lack formal recognition. Their defining characteristics remain unexplored. We aim to establish a formal definition of these scams, aiding in the identification of similar patterns in \acro LPs and facilitating their distinction from legitimate ones.
    
    \item \textbf{RQ2.} \textit{Can we build and evaluate a rule-based detector to identify \acro scams?} 
    Using the identified characteristics of \acro scams, we aim to design a rule-based heuristic to detect \acro LPs on Ethereum. This heuristic will be evaluated in conjunction with external data from legitimate LPs. Additionally, we aim to gain deeper insights into \acro scams by analyzing cases with unexpected outcomes (i.e., false positives, false negatives).
    
    %Using the common structures and characteristics in identified from scams, we designed a rule-based heuristic to discover the \acro LPs on Ethereum. This part focuses on developing and evaluating the heuristic using our initial ground truth dataset from RQ1 and the external legitimate LP. From the evaluation perspective, we also aim to uncover deeper insights into the \acro scams by analyzing instances with unexpected heuristic flagging results (i.e., false positive and negative cases). 
    
    \item \textbf{RQ3.} \textit{Can we use the heuristic to automatically detect \acro scams at a large scale? }
    Building on RQ2, we aim to deploy the heuristic at scale to identify and label SLID scams across the six largest Ethereum DEXs. However, the heuristic’s reliance on a long historical window introduces detection latency, and its inherently fuzzy design may fail to capture borderline cases. To address these limitations, we propose developing a proactive detection framework that enables earlier warning signals and improves coverage of ambiguous or borderline SLID instances.
    % Based on RQ2, we aim to apply the heuristic at scale to identify and label all \acro scams across the six largest Ethereum DEXs. However, the heuristic's dependency on a long historical range of transaction data introduces latency in flagging \acro scams, and the heuristic's fuzzy design might fail to capture the borderline cases. To address this limitation, we aim to develop a more proactive method capable of providing early warnings and broader SLID catch on the borderline SLID cases.
\end{packeditemize}

We first established our initial SLID-like dataset by identifying community-reported DeFi scam cases that were uncategorized and exhibited characteristics similar to our research motivation. This process resulted in the identification of 71 LPs that matched these descriptions. We then conducted a pilot study to analyze common scam behaviors shared by this initial SLID-like dataset (\S\ref{sec:rq1}). Based on the insights extracted from this analysis, we designed an efficient rule-based heuristic capable of identifying SLID LPs by examining their DEX activity records and smart contract security features. This heuristic was validated using legitimate LPs, with evaluation outcomes discussed in detail (\S\ref{sec:rq2}).

Next, we applied our rule-based heuristic to 319,166 Ethereum LPs across the six largest DEXs on Ethereum (\S\ref{sec:ml}). This resulted in the flagging of 3,117 SLID scams, which collectively generated a profit of \$103.2M on these DEXs as of July 13, 2024. This analysis introduced the first-ever ground truth dataset of confirmed SLID scams. We then analyzed the flagged instances through three key dimensions: vitality, trend analysis, and profit, providing deeper insights into the tactics and the overall impact of SLID scams. Leveraging this discovery of SLID ground truth, we applied three machine learning (ML) algorithms (\S\ref{subsec:ML}) of Random Forest, XGBoost, and Logistic Regression for early detection of SLID scams, capable of detect the borderline cases that have been left from our SLID fuzzy heuristic and identifying potential scams with such a shorter required DEX activity history that can detect \acro scam 4.77 times faster than the heuristic while maintaining high detection F1-Score of 0.93. 

In summary, we made the following key contributions:
\begin{packeditemize}
    \item We presented the first formal definition of the SLID scam in DEX environments. We conducted an efficient pilot study to explore its characteristics and design a rule-based heuristic. We then evaluate the heuristic and discuss insights behind the different true and false positive detections.

    \item We applied our heuristic to six major DEXs on the Ethereum network, identifying 3,117 LPs exhibiting SLID behaviors (with responsible disclosure). We produced the first ground truth dataset for SLID scams, publicly released at \url{https://anonymous.4open.science/r/SLID-sourcecode-41AD}, and enabled analysis across vitality, profit, and trend metrics. We also responsibly disclosed our findings to affected DEX teams and made all our coding resources used in this study publicly available \cite{slidsource}.
    
    \item To enable early warning and discover borderline cases outside of the heuristic scope, we further applied ML algorithms via our ground truth dataset. We successfully reduced detection time by 4.77x while maintaining a 0.93 F1-score.
\end{packeditemize}

\begin{takeaway}
\noindent\textbf{Responsible disclosure.} We have reported our findings to the specified DEXs, i.e., Uniswap, Curve, Balancer, Bancor, and PancakeSwap, to raise their awareness of such scams.
\end{takeaway}

\section{Technical Warm-Ups}\label{sec:bck}

\noindent\textbf{Ethereum tokens.}
Ethereum~\cite{eth,wood2014ethereum} is the 2nd leading blockchain after Bitcoin \cite{btcintro}. Ethereum introduced \textit{smart contracts} and enabled the automatic execution of code. Users can create tradeable digital assets on-chain, known as \textit{tokens}. Those tokens are fungible (adhering to ERC20~\cite{erc20}) and serve multiple purposes, including fuelling decentralised applications (DApps), acting as stakes, or representing ownership. 

Ethereum accounts are tied to unique key pairs in two types~\cite{wang2023account}: (\textit{i}) \textit{externally owned accounts} (EOAs) store tokens and facilitate interactions with other accounts; (\textit{ii}) \textit{contract accounts} are associated with smart contracts to provide specialized on-chain services.

\smallskip
\noindent\textbf{DEXs and liquidity pool.}~\label{dexbackground}
Uniswap \cite{uniswap0}, the unicorn DEX on Ethereum, is the vital DApp that has significantly contributed to the growth of the DeFi ecosystem. Their AMM infrastructure \cite{dexamm} enables users to list their newly created tokens for trading by deploying a liquidity pool (LP). A LP is a smart contract on DEXs that holds two (or more) tokens, typically including at least one well-known token with independent value (e.g., ETH, USDT, or USDC), referred to as the \textit{base token}. The other token is another base token or a newly introduced token, known as the \textit{paired token}. 

When initializing a LP, the owner must deposit a specific amount of both tokens. The value of this LP is determined by the total USD value of the base token it holds, while the price of the paired token is positively correlated to the LP's overall value, following the provided AMM equations \cite{ammformula, uniswapdexpriceexplain}. Trading the base token for the paired token is termed a {\it buy}, while trading in the opposite direction is a {\it sell}. These swaps impact the LP's dynamics: \textit{buy} increases the price of the traded token as more base tokens flow in, whereas \textit{sell} reduces prices as tokens are withdrawn. 

Any user can contribute to the LP's liquidity via \textit{deposit} with a certain amount of both tokens to earn incentives later or through \textit{withdraw} to remove their previously provided liquidity.  All activities are publicly recorded on-chain and are referred to as DEX orders or DEX activities of that LP.

When it is proven that the price of the paired token is correlated with the liquidity of the base token in the LP, the owner can guarantee financial safety when pulling the scam and always exit at a higher price to get profit (since when the trader comes, the liquidity of the base token first increases and leads to the increment of the price of the paired token, see more details at Appendix~\ref{appendixproof}).

Legitimate LPs conduct the action of burning (i.e., permanently locking) initial LP tokens after deployment. This process removes the LP owner's ability to control the liquidity, preventing them from taking unethical actions, such as draining the LP. Burning LP tokens is widely regarded as an industry standard for establishing trustworthiness and transparency in the DeFi space \cite{sokdefi,jiang2023decentralized,uniswap0,lpburn}.

\smallskip
\noindent\textbf{Rug pull.}~\label{scambackground}
The most common DEX scam leveraging LPs is {\it rug pull} \cite{rugpull}. The scammer creates (so that owns) the scam LP and withdraws all the base tokens in the LP later. There are two basic operations through which scammers extract profits: 
\begin{packeditemize}
    \item \textit{Sell}: first, sell a large inflated amount of paired tokens (they owned) in exchange for all LP's base tokens; 
    \item \textit{Withdraw}: withdraw the liquidity they initially provided, ``pulling'' out all of the base tokens from the LP.
\end{packeditemize}
When scammers engage in sell or withdraw activities and the value of the retrieved assets exceeds their initial investment, those excess amounts are defined as the scammer's \textit{realized profits}. Correspondingly, \textit{unrealized profit} represents the remaining value of assets within the scam LP that still belong to the scammer but have not yet been converted into \textit{realized profits} via sell or withdraw actions.

Those profit-taking actions caused a sharp drop in the token’s price and trading volume. Investors incur immediate financial losses. New investors lose interest in the scam LP. These patterns serve as signals for detecting rug pulls.

\smallskip
\noindent\textbf{Honeypot.}
Another typical scam is {\it Honeypot} \cite{honeypot}. The scammer maliciously designs the token's smart contract and allows users to easily purchase tokens but hardly sells them. The contract may include restrictions such as preventing sell transactions, allowing withdrawals only for whitelisted addresses, or imposing exorbitant transaction fees that make it impossible to retrieve funds. The scammer then exploits LPs through sell/withdraw actions, similar to the mechanics of rug pull.

We defer more related work to Appendix~\ref{sec:rw}.

\begin{takeaway1}
Rug pull and Honeypot share common characteristics: (i) LP tokens (LPTs), which represent the initial liquidity~\cite{bartoletti2021sok}, are not burned, and (ii) the ownership of the paired token remains entirely monopolized by the LP creator. 
\end{takeaway1}

Without implementing an ethical safeguard (i.e., burning initial LPTs), liquidity providers, including those involved in rug pulls, honeypots, and SLID scams, confront no financial risk (proven in Appendix~\ref{appendixproof}) for engaging in fraudulent activities.

% While honeypot scams require the scammer to alter the token's smart contract, they can be detected since its deployment. In contrast, the \acro scam leaves no traces of smart contract interference, preventing them from being identified early. The \acro scam shares similarities with the rug pull scam during the operational phase, as the scammer is also the LP owner and deployed the scam liquidity LP. This newly identified scam profits by draining the LP's base token through \textit{sell}/\textit{withdraw} actions. However, unlike rug pull, which drains liquidity quickly, the \acro scammer operates at a gradual and slow pace, executing multiple profit-taking actions with small proportions over an extended period. 
% This behavior allows the \acro scammer to remain undetected by both scam detectors and users, as it does not cause significant fluctuations in the token's price or trading volume.
% As a result, the \acro scam appears to be a normal and safe investment, attracting new investors and causing the liquidity LP's value to grow over time. This ultimately leads to a more favorable profit ratio for the scammer during their future profit-taking activities.

%======================================
\section{Motivating Examples}
\label{subsec:motivation}
%======================================

\subsection{Case Study of Initial SLID Liquidity Pool}
% In conducting this study, we anonymized the collected data, including token names, addresses, and liquidity LP details, to ensure security and privacy. This section discusses the behavior of token A with token address "0x4e84...a17c8", created in August 2021, which serves as a motivational example for our study. This token was launched with a massive marketing campaign highlighting its high potential for generating profits for investors. We have identified the liquidity LP of A as a \acro due to the deceptive actions of its owners, who slowly exploited the liquidity LP through a mixed sequence of token buying and selling on Uniswap while maintaining a stable price and trading volume during the initial months after its launch. The gradual profit-taking persisted for an extended period before the token owner executed a significant rug pull a year later. This section briefly describes the sequence of actions taken by the owners in this cryptocurrency's \acro scam.
% \begin{figure}[h]
%   \centering
%     \includegraphics[width=\linewidth]{graphs/SLID.png}

%     \caption{\textbf{The workflow of our SLID scam:} \ding{172} create a liquidity LP; \ding{173} add tokens (base \& paired); \ding{174} avoid burning initial shares/tokens; \ding{175} marketing campaign; \ding{176} invest funds and buy tokens; \ding{177} normal activities (buy orders) \ding{178} slowly withdraw/sell funds \ding{179} liquidity drop; \ding{180} money loss.}
%  \label{fig:add-mempool}

% \end{figure}
We provide a typical case study. We discuss the behavior of Token~\textcolor{teal}{$\hat{A}$} (here, our collected data, including token names, addresses, and LP details, are anonymized for ethical reasons) with token address ``0x4e84...a17c8'', created in August 2021, which serves as a motivating example for our study. This token was launched with a massive marketing campaign highlighting its high potential for generating profits for investors, and later got reported by the public (details in \S\ref{subsec:groundtruth1}) for the owner's fraudulent selling activities. 

We identified the LP of \textcolor{teal}{$\hat{A}$} as a \acro scam due to the deceptive actions of its owners. They slowly exploited the LP through a mixed sequence of token buying and selling on Uniswap while maintaining a stable price and trading volume during the initial months after its launch. The gradual profit-taking persisted for an extended period before the token owner executed a significant withdraw a year later. We briefly describe the sequence of actions taken by the owners in this cryptocurrency's \acro scam.

\smallskip
\noindent\textbf{\ding{202} Attracting victims.}
To rapidly attract users, \textcolor{teal}{$\hat{A}$}'s advertising campaign targeted buyers by promising substantial token profits if purchased early. The owner engaged the community by promoting the token via their X account (formerly Twitter). They attracted 78.5K followers and collaborated with 20+ well-known crypto influencers. The campaign drew 5,326 investors to purchase \textcolor{teal}{$\hat{A}$} on Uniswap within the first 70 days from the token's LP deployment.

\begin{figure}[t]
  \centering
    \includegraphics[width=\linewidth]{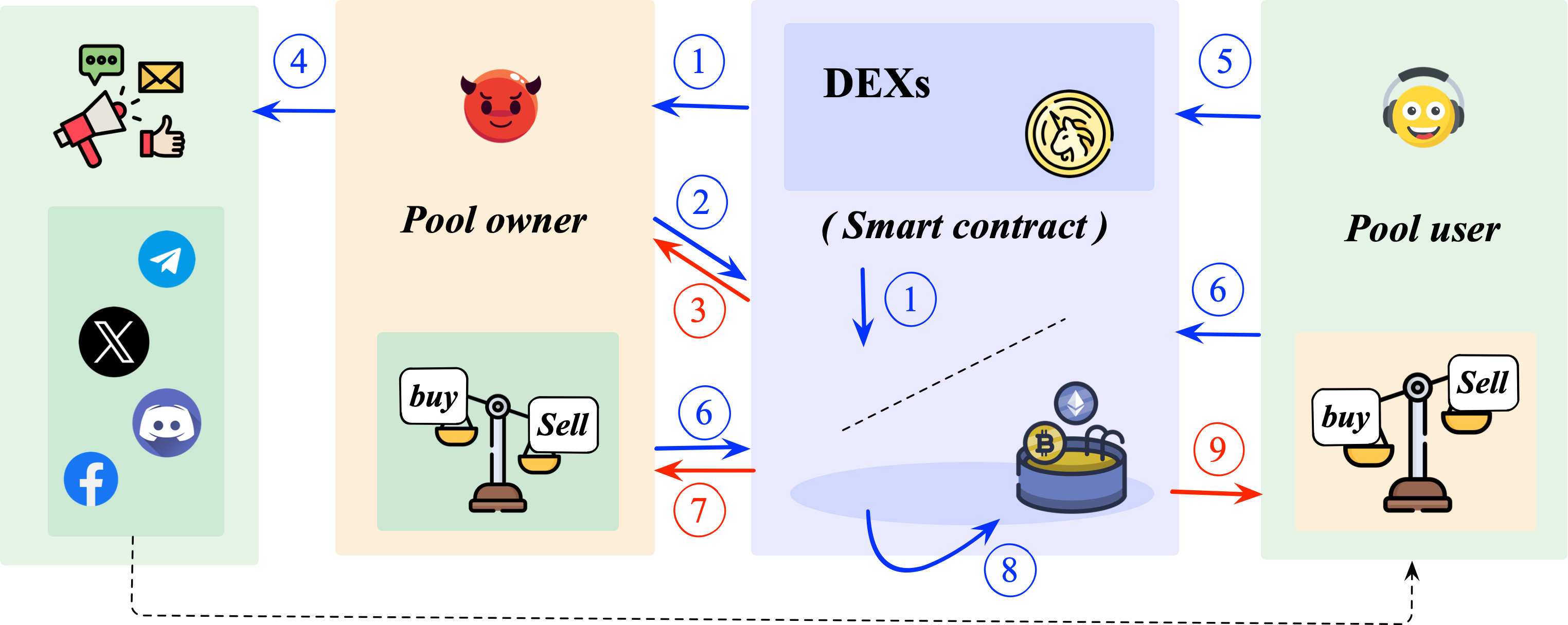}
    \caption{\textbf{SLID workflow:} \ding{172} create a liquidity pool; \ding{173} add tokens (base \& paired); \textcolor{red}{\ding{174}} avoid burning initial shares; {\ding{175}} marketing campaign; \ding{176} invest funds and buy tokens; \ding{177} normal activities (buy orders);  \textcolor{red}{\ding{178}} slowly withdraw/sell funds; \ding{179} liquidity drop;  \textcolor{red}{\ding{180}} money loss.}
 \label{fig:add-mempool}
 \vspace{-0.2in}
\end{figure}

\smallskip
\noindent\textbf{\ding{203} Maintaining the LP and slowly taking profit.}~\label{subsubsec:motivationchar}
The LP was deployed on Aug. 4th, 2021, with an initial owner liquidity deposit of \$19K. However, the LP for \textcolor{teal}{$\hat{A}$} was deployed without burning the owner's initial tokens. This creates a persistent risk, as the owner of \textcolor{teal}{$\hat{A}$} can drain the LP and seize all funds invested by users to purchase \textcolor{teal}{$\hat{A}$} at any time. This is precisely what occurs in rug pull and honeypot scenarios (cf. \S\ref{sec:bck}).

\begin{figure*}
  \centering
  \includegraphics[width=0.99\linewidth]{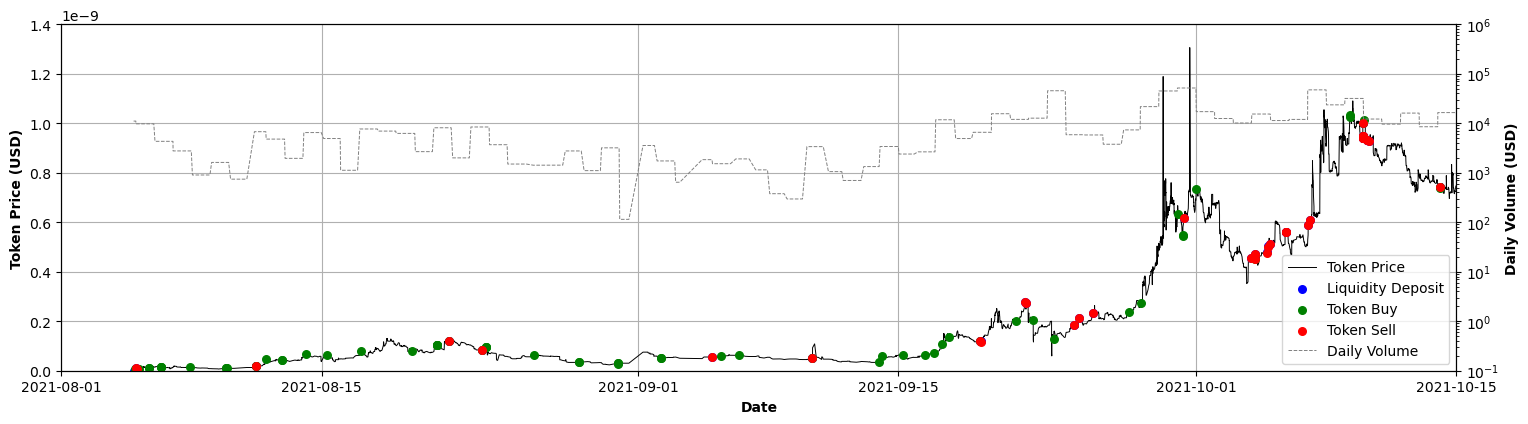}
  \caption{70-day \textbf{scammer activities} on Uniswap with token price and trading volume.}
  \label{fig:luffy}
\end{figure*}

We observed \textcolor{teal}{$\hat{A}$}'s owner DEX activities (Figure~\ref{fig:luffy}) over 70 days after deployment, including price and trading volume. During this period, the owner was engaged in a sequence of mixed activities, including buying and selling. Although these actions seem legitimate, our evaluation reveals that the selling amount \textbf{consistently exceeded} the buying amount (in \textit{paired tokens}). Specifically, the total selling volume was \textbf{ten} times larger, with the owner purchasing only $2.38 \times 10^{15}$ tokens while selling $3.21 \times 10^{16}$ tokens of \textcolor{teal}{$\hat{A}$} in total. This indicates that the owner possesses a significant number of tokens and attempts to incorporate this inflated amount into their selling orders. \textcolor{teal}{$\hat{A}$}'s owner executed 139 buy and 136 sell actions, but with a \textbf{double} volume gap that was worth \$783K and \$1.4M, respectively. This resulted in a huge profit for the LP owner of \$707K — double the initial investment.
In contrast, 5,326 investors committed with 10,374 buys, 3,898 deposits, and 7,322 sells, but \textbf{only} 2 withdraw activities, with a total inflow of \$1.35M (in base token) to the LP. 
The mix of buy and sell orders created the appearance of legitimacy in the token's transaction history, helping to avoid the token price crash that would occur if they sold all at once (i.e., 1-day rug pull~\cite{spammer}).

\smallskip
\noindent\textbf{\ding{204} Repeating the scam.}
While maintaining the appearance of token legitimacy, \textcolor{teal}{$\hat{A}$}'s owner continued to sell only a small portion of the tokens compared to the overall LP value. The smallest and largest sell order values account for only 1\% and 16\% of the LP's value. 
By carefully mixing buy and sell orders and controlling the selling amounts, the built-in scam detectors on DEXs are \textbf{unable} to identify this token as a scam. As of July 13, 2024, the scammer still maintains \$91K worth of liquidity in the LP and has remained inactive since their last recorded sell order on July 31, 2022.

Thus, despite the token's stable behavior, a series of {\it buy} and {\it sell} actions (with smaller amounts) occurred at a slow pace, gradually adding an inflated amount of paired tokens (\S\ref{sec:definition}) to each sell order from the LP owner's account.
This scam strategy can easily avoid investor attention while enabling tokens to remain active and attract additional investments over an extended period. The transaction history of \textcolor{teal}{$\hat{A}$}'s LP on Uniswap in the analyzed period is detailed in \cite{motivdata}.

% \begin{figure}[t]
%     \centering
%     \includegraphics[width=\linewidth]{graphs/SLID-V3.png}

%     \caption{SLID scam\qw{1. increase fontsize, too small now 2. reorganize the boxes a bit, to reduce blank space, fit it into a full rectangular (example Fig2 https://arxiv.org/pdf/2308.10422) 3. as there are 8 steps, the index should be called/cued in the corespoining paragraphs like xxxx (step 1 in Fig1) 4. maybe timesNewRoman peforms better look in font [all same to Fig5]} \nasrin{have done 1, 2, and 4. Where have we referred to this figure?}}
%     \label{fig:add-mempool}

\subsection{More LPs Exhibiting SLID Patterns}

In addition to the motivation example, we present more LPs, i.e., \textcolor{teal}{$\hat{B}$} and \textcolor{teal}{$\hat{C}$}, exhibiting similar operational patterns. All those projects receive community public reports (\S\ref{subsec:groundtruth1}).

\smallskip
\noindent\textbf{LP of token \textcolor{teal}{$\hat{B}$}.} LP \textcolor{teal}{$\hat{B}$} (with paired token address ``0x59...e9CD") followed a suspiciously similar lifecycle, beginning with the initial deployment of an LP on 12 Feb 2024 without burning LP tokens, granting persistent unilateral control to its creators. Our analysis of transaction records revealed that the owner obtained a huge amount of inflated paired tokens minted outside and continuously sold them into the LP from 12 February to 17 April in a moderate portion. The owner incrementally extracted investor funds of \$2660.9 after the first 2 months of the LP's deployment. 

Notably, up to today, token \textcolor{teal}{$\hat{B}$}'s trading activity in the same LP showed frequent but minor liquidity withdrawals interspersed with small buy and sell orders, similar to \textcolor{teal}{$\hat{A}$}.
The only difference is that \textcolor{teal}{$\hat{B}$} does not run a large marketing campaign. Despite attracting fewer buyers, the owner still generated illicit profits.

\smallskip
\noindent\textbf{LP of token \textcolor{teal}{$\hat{C}$}.} LP \textcolor{teal}{$\hat{C}$} (token address ``0xbA...9ccE") also exhibited trading patterns indicative of SLID behaviors. The LP was deployed on 7 April 2024. The owner maintained control without burning LP tokens, possessed inflated tokens externally, and gradually sold them with 38 sell orders from April 9 to April 26, 2024. Meanwhile, LP \textcolor{teal}{$\hat{C}$} displayed frequent, small-scale liquidity withdrawals combined with sporadic minor transactions. The LP only employed minimal promotional activity from X's influencer. Unlike other examples, this scam resulted in a realized profit of \$200 for the owner. The realized profit appeared to be smaller compared to other scams during the same period. \textcolor{teal}{$\hat{C}$} transaction history still illustrates the core SLID behavior (more details in \S\ref{sec:definition}). 

All the transactions involved in the motivation examples analysis of \textcolor{teal}{$\hat{A}$}, \textcolor{teal}{$\hat{B}$} and \textcolor{teal}{$\hat{C}$} are provided in \cite{motivdata}.

\begin{takeaway1}
\noindent\textbf{Side finding:} SLID scams strategically exploit incremental deception to evade immediate detection. They are not isolated but a broader fraudulent pattern across DeFi. 
\end{takeaway1}

\section{Common Features of \acro Scams (RQ1)}\label{sec:rq1}

Based on examples, we developed a multi-layered filtering mechanism~(\S\ref{subsec:groundtruth1}) to initiate our pilot study and find the scams with similar features (i.e., SLID-like LPs,~\S\ref{subsec:char}).

\subsection{Identification of \acro LPs}\label{subsec:groundtruth1}

We began by collecting publicly available DeFi scam reports from a broad range of sources dating back to 2022. These sources of reports included well-known platforms such as Chainabuse \cite{chainabuse}, De.fi \cite{dedotfi}, Chainalysis \cite{chainana}, and prominent scam warning accounts on X (formerly Twitter), including ZachXBT \cite{zach} and Pocket Universe \cite{pocket}. From this collection, we identified \textbf{921} reported LPs that formed our initial dataset (can be accessed in \cite{initialgroundtruthdata}).

We then developed and applied a four-layer filtering system to isolate LPs that exhibited SLID-like characteristics:

\begin{packeditemize}
\item \textbf{Owner profit-checking layer.} The layer computes the realized profit of the LP owner based on their profit-taking action from trading history and then excludes LPs in which the owner has a negative realized profit (Appendix~\ref{appendixeq1}). The layer ensures accuracy by filtering out the falsified reports where tokens were not fraudulent but reported as a scam. As a result, \textbf{898} LPs bring profits to the owner out of 921 reported LPs, and 23 LPs were excluded. 

\item \textbf{Honeypot layer.} The detection against honeypot is straightforward in LP's deployment phase. We filtered out instances that contain honeypot vulnerability by utilising GoPlus \cite{goplus} and De.Fi \cite{dedotfi} services. We excluded 139 honeypot LPs out of 898 LPs. The remaining \textbf{759} LPs were eligible.

\item \textbf{Mislabelled LPs filtering layer.} This layer processes the LP instances and cross-references them with the rug pull detection methods (i.e., \cite{rugpull,spammer}), to determine if any of the collected scams were rug pull and filters them out from the SLID-like LPs. Out of the 759 input instances, 665 instances were recognized as rug pull and filtered out, while \textbf{94} LPs moved to the next layer.

\item \textbf{Owner action layer.} Amongst the 94 remaining instances, we recognized that they may contain various other undetermined DeFi frauds that are irrelevant to SLID. We thus selected LPs (i) without the initial LP burned and (ii) having at least three DEX activities conducted by owners, including deposit, withdraw, buy, or sell (aligning with our observations in \S\ref{subsubsec:motivationchar}). As a result, we excluded 23 LPs and identified \textbf{71} LPs that passed all four layers.

\begin{takeaway1}
With our filtering mechanism, we constructed our initial SLID-like LP dataset, consisting of 71 (out of 921) pools.
\end{takeaway1}
% As a result, we excluded 23 LPs and identified \textbf{71} LPs that passed all four layers as our \textcolor{red}{\textbf{SLID-like LPs dataset} for RQ1. 

\end{packeditemize}

\subsection{Finding and Refining Features from SLID-like LPs by Comparing with Rug Pulls}\label{subsec:char}

To identify distinguishing features, we conducted an in-depth analysis and benchmarked the 71 filtered LPs against rug pull tokens that had been excluded \textit{mislabeled LPs filtering layer} (\S\ref{subsec:groundtruth1}). These rug pulls were reintroduced as a behavioral baseline for comparison, as they exhibit two known fraudulent profit-taking strategies: (\textit{i}) inflated selling and (\textit{ii}) initial liquidity pool (LP) withdrawal.

% To identify the distinguishing features, we conducted an in-depth analysis on the 71 filtered LPs. Here, we included rug pull tokens that had been excluded by \textit{mislabeled LPs filtering layer} (\S\ref{subsec:groundtruth1}). Rug pulls were selected as a behavioral benchmark for comparison due to their shared use of two fraudulent profit-taking strategies: (i) inflated selling and (ii) initial LP withdrawal. 

Our analyses identify three defining properties (\textbf{P1}, \textbf{P2}, \textbf{P3}) commonly observed in SLID LPs.

\begin{figure}[h]
    \centering
    \includegraphics[width=0.9\linewidth]{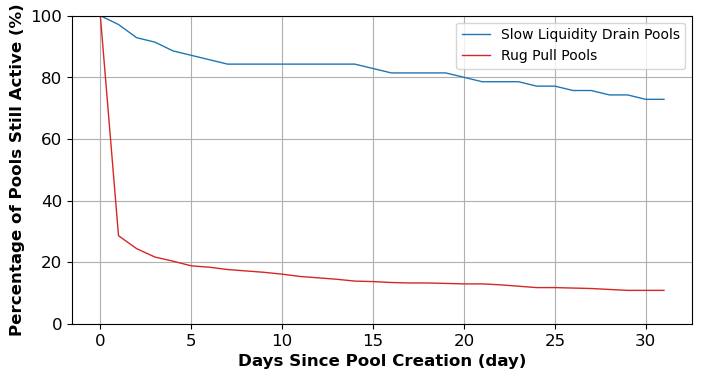}
    \caption{30-day \textbf{liveliness} of rug pull and \acro LPs.}
    \label{fig:alive}
\end{figure}

\smallskip
\subsubsection{\underline{Post-month liveliness analysis}}\label{subsubsec:postmonthliveliness} Liveness is one key metric to detect scam tokens~\cite{spammer}\cite{sadponzi}\cite{nftrug}. Nearly 60\% of the rug pull LPs ended up ``inactive'' after 1 day of deployment~\cite{spammer}. 

We observed LP lifetimes for both scam types over an extended period of one month, called \textit{post-month liveliness}. We also compared the average changes in user activity for LPs that remain ``active'' after the first month to distinguish \acro scams from rug pulls. 

Our results revealed that 72.85\% of SLID-like LPs remain active after the first month of deployment, whereas only 10.8\% of rug pull LPs persist beyond this threshold. On the first day of the second month, trading activities for the remaining active \acro LPs were stable, showing an average reduction of 38.16\% compared to the LP's previous peak day with the highest recorded activity count. In contrast, the remaining rug pull LPs experienced a sharp decrease (95.82\%) in recorded activities on the first day of the second month. 

The findings led to our conclusion: 
\begin{packeditemize}
    \item[\textbf{P1.}] \textit{The \acro LPs are more adept at remaining undetected due to their prolonged liveliness. They remain active over a longer duration compared to rug pull despite using similar LP initiation and profit-taking mechanisms, making it nearly impossible to detect based solely on liveliness status.}
\end{packeditemize}

%While various cryptocurrency fraud studies (e.g., \cite{spammer} on rug pull, \cite{sadponzi} on Ponzi schemes, \cite{nftrug} on NFT rug pull) rely on the token's liveliness as a vital factor for identifying scams, this metric appears to be ineffective in the \acro case due to its strong survival trait.

% \begin{itemize}
%     \item[\textbf{C1}] \textit{The \acro LPs are better at remaining hidden due to its long liveliness. This scam type prefers a longer period than rug pull with the same profit-taking mechanism, making it nearly impossible to detect relying on liveliness status.} 
% \end{itemize}

% \subsubsection{Scammer Activity Analysis}\label{subsubsec:activity}
% For better observation of the owner's financial behavior and scam motivation for both fraud types, we analyzed the scammer activity pattern for all targeted LPs after the first month. Since the scammer of these LPs is also the owner and creator of it, there are two mechanisms that both rug pull and \textit{slow liquidity drain} commonly use for profit-taking. They can either withdraw the LP when the victim buys the tokens (and enlarge the LP's value) or sell the inflated amount of tokens that they have outside to drain the LP.

%We analyzed the scammer's activity patterns in all targeted LPs after the first month to learn the owner's financial behaviors and motivations behind both fraud types. Given that the scammer is also the owner of these LPs, after victims buy paired tokens and send base tokens in the LP (LP value thus increased), 

\subsubsection{\underline{Scammer activity analysis}}\label{subsubsec:activity} For both rug pulls and \acro scams, scammers were LPs owners and relied on two strategies: (\textit{i}) withdrawing liquidity from the LP and (\textit{ii}) selling the inflated tokens they held externally to drain the LP. We then analyzed the scammer's activity patterns

Our results show that rug pull scammers prioritize \textbf{immediate} profit-taking, with 99.7\% (663/665) of LPs being entirely drained within the first day, and 100\% drained within the first three days. The LP owners, who are also scammers, perform 75.38 DEX activities to their LP and 15.79 profit-taking actions on average during the first month. Among these profit-taking actions, each LP includes a \textbf{single}, \textbf{large-amount} profit-taking transaction (either selling or withdrawing) that constitutes over 95\% of the total LP value. On average, those scammers extract 99.33\% of the LP’s value after the first month, leaving only 0.67\% as unrealized profit. 

In contrast to rug pulls, scammers in SLID LPs operate more gradually, conducting \textbf{multiple} profit-taking activities with \textbf{smaller} proportions. None of the SLID LPs were completely drained within the first three days after deployment, unlike 100\% of rug pull LPs, which were fully drained in that same timeframe. On average, SLID scammers executed 996.59 total orders during the LP’s first month, over 10 times more than rug pull LP owners. They also carried out a mean of 423 profit-taking actions, approximately 26 times more frequent than rug pull scammers. Additionally, SLID scammers keep their profit-taking within a narrower range, with drained values varying from 7.39\% to 42.93\% of the LP’s total value. This controlled approach can evade detection by standard rug pull detectors.

% In contrast, scammers in \acro LPs use contrasting strategies, operating more gradually and performing \textbf{multiple} profit-taking activities with \textbf{smaller} proportions. We observed that none of the \acro LPs were completely drained within the first three days after deployment, in stark contrast to the 100\% of all LPs that were totally drained in rug pull scams during the same period. On average, \acro scammers executed 996.59 total orders during the LP's first month, which was over 10 times more than the rug pull LP owners. \acro owners had a mean of 423 profit-taking actions, approximately 26 times more frequent than rug pull scammers. Additionally, \acro scammers control the proportion of profit-taking orders in a smaller range. The amount of drained value from their profit-taking varies from a minimum of 7.39\% to a maximum of 42.93\% of the LP's total value on average. This controlled proportion enables \acro scammers to evade detection by rug pull detectors.

\begin{figure}[h]
    \centering
    \includegraphics[width=.99\linewidth]{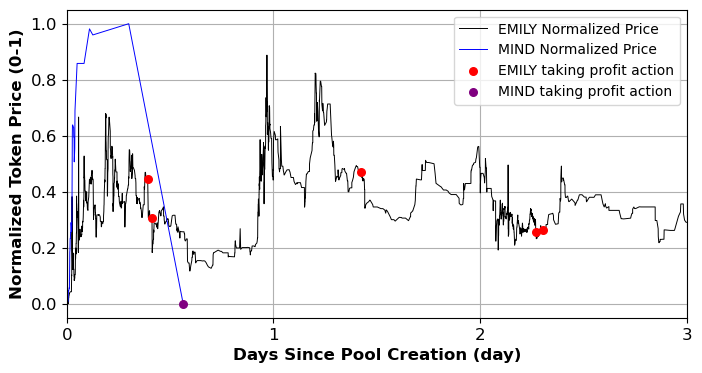}
    \vspace{-0.03in}
    \caption{Token prices of EMILY (SLID-like) and MIND (rug pull) were affected by scammer's profit-taking activities in the first 3 days.}
    \label{fig:pricecompare}
\end{figure}

To better demonstrate differences, we further provided two observed example tokens (Figure~\ref{fig:pricecompare}): MIND~\cite{mind} (rug pull) and EMILY \cite{emily} (SLID-like). While a single profit-draining action by the MIND owner completely emptied LP and reduced the token's price to zero, the EMILY owner conducted multiple small-proportion exits that do not heavily affect the token's price. Although both scam types initiate fraud similarly by creating LPs and profiting through draining them via sell or withdraw orders, we observed that variations in execution duration and order proportions lead to distinct strategies for each scam type. We thus conclude that
\begin{packeditemize}
    \item[\textbf{P2.}] \textit{Although \acro scams employ a mechanism similar to that of rug pull, they pursue an opposite strategy characterised by a gradual approach, avoiding recognition. Scammers prioritise multiple small-proportion and slow-paced profit extraction activities, making it difficult to detect, either through observation or by using detectors.} 
\end{packeditemize}

\subsubsection{\underline{Profit analysis}}\label{subsubsec:proft}
The \acro strategy is intuitively expected to yield a better profit compared to rug pull by allowing scammers to conceal activities while gradually attracting more victims.
We validate this speculation via a profit-changing analysis between the two scam types. Technically, we focus on both \textit{realized} and \textit{unrealized} profits of the LP owner. 

% Our results showed that the realized profits \sout{(the actual profits sold or withdrawn by scammers from the LP), refer to appendix??} significantly differ between the two scam types. 

Our results showed that the realized profits (explained in \S\ref{sec:bck}). \acro scams demonstrate a superior profit margin: in the first month, while rug pull scammers achieve an average realized profit of 7.5 times their investment through a single, short-term draining action, \acro LPs attain a massively higher profit margin of 10.3 times their investment through multiple small draining activities. As for unrealized profits, \acro scams also perform a notable gap in profit advantage. Rug pull scammers have nearly zero future profit while completely emptying their LPs, retaining only an average of 0.67\% of LP value. In contrast, the \acro LP owner's drainable assets in the LP after the first month retain, on average, 156\% worth of their initial investment. This allows the scammer to either fully drain the unrealized profit, resulting in an additional 1.56 times profit on top of the realized profit (already taken), or to refrain from withdrawing and let LP continue to grow. 

We found that, by performing a vast number of small-proportion profit-taking orders (\S\ref{subsubsec:activity}), \acro scammers manage to attract more victims, resulting in a larger netflow of funds poured into LPs. Although they only take part of their possible profit in the first month, these scammers already achieve better profit margins than rug pull scammers. Furthermore, a significant portion of assets that remain in the \acro LPs retain substantial profit-scaling potential as the LP continues to grow, entrapping more future buyers. 

Therefore, we conclude that,
\begin{packeditemize}
    \item[\textbf{P3}] \textit{\acro scams offer a higher profit margin and sustainability, providing both significant mid-term profits and future profit-scaling potential if undetected. Compared to rug pull, \acro scammers can achieve a substantial growth rate in LP value from their initial investment, making it a more lucrative and enduring scam approach.}
\end{packeditemize}

\vspace{-3pt}
\begin{takeaway}
    \noindent\textbf{Findings of RQ1.}
     SLID scams execute numerous small, profit-taking actions over extended periods, interleaved with routine LP owner activities to obscure malicious intent. This behavior enables scammers to remain active and gain profit while evading detection systems tuned for abrupt fraud.     
\end{takeaway}

% \begin{takeaway}
% \noindent\textbf{Findings of RQ1:}
%         An \acro scam employs a gradual tactic involving multiple small-proportion profit-taking actions, interspersed with other owner activities to create misdirection. This strategy enables the scam to evade detection over an extended period, maintaining LP activity while steadily increasing profit margins. 
% \end{takeaway}

%These identified patterns offer valuable insights to uncover previously undetected \acro LPs, thereby later improving the accuracy and comprehensiveness of our ground truth dataset.
\section{Rule-based  \acro Detection (RQ2)}\label{sec:rq2}

Based on our empirical insights (\S\ref{sec:rq1}), we step further by formally defining SLID fraudulent (\S\ref{sec:definition}) actions and introducing our heuristic (\S\ref{subsec:heuristic}).

%=========================================
\subsection{SLID with A Formal Understanding}
\label{sec:definition}

\noindent\textbf{System model.}
We consider a DEX where an LP $\mathcal{P}$ (liquidity pool) allows trading between:
\begin{itemize}
    \item A \textbf{base token}, denoted $B$ (e.g., ETH, USDC), and
    \item A \textbf{paired token}, denoted $P$ (typically a new or obscure token).
\end{itemize}
 
There are two main participants:
\begin{itemize}
    \item \textbf{Owner} ($O$): deploys the LP and adds initial liquidity.
    \item \textbf{Investors}: interact with the LP through swaps, deposits, and withdrawals.
\end{itemize}

\smallskip
We define the following terms:
\vspace{-0.3em}

\begin{itemize}
    \item $t_0$: time of LP deployment.
    \item $B_0, P_0$: initial deposited amounts of base and paired tokens.
    \item $p^t$: price of token $P$ in terms of $B$ at time $t$.
    \item $B^t$: base token balance in the pool at time $t$.
    \item $V^t$: trading volume at time $t$.
    \item $X^t$, $Y^t$: total amount of $P$ bought/sold by users up to time $t$.
    \item $W^t$: cumulative amount of $B$ withdrawn by $O$ by time $t$.
\end{itemize}

We assume the owner:
\begin{itemize}
    \item Retains full control over the initial LP tokens (denoted $\texttt{LPT}_{\text{init}}$).
    \item Possesses an amount of inflated paired tokens $P$.
\end{itemize}

%\subsection{Fraudulent Actions}
%\label{inflatedsell}

We identify three core fraudulent actions (rooted in \S\ref{subsec:motivation}).

\begin{center}
\fbox{%
\begin{minipage}{0.9\linewidth}

\begin{itemize}
 
\item[\ding{172}] \textbf{Retained LP Token Control.}
The owner does not burn initial LP tokens:
\[
\texttt{burn}(\texttt{LPT}_{\text{init}}) = \text{false}
\]
This permits future liquidity withdrawals without investor approval.
 
\item[\ding{173}] \textbf{Inflated token selling.}
The owner sells their private inflated supply of $P$ over time via small-volume trades:
\[
\mathcal{S} = \{ (p^{i}, q^{i}) \}_{i=1}^{n}, 
\]
where each $q^i$ is a single trade of an owner's $P$ token at price $p^i$ (at time $t_i$) satisfying:
 \begin{align*}
  q^i &< \delta \cdot B^{t_i}, \quad (\text{small relative to pool size}) \\
|p^{i+1} - p^i| &\approx 0 \quad (\text{stable price})
\end{align*}

with $\delta \ll 1$.
 
\item[\ding{174}] \textbf{Incremental liquidity withdrawals.}
The owner performs multiple small withdrawals of base tokens:
\[
\mathcal{W} = \{ w^j \}_{j=1}^m 
\]
such that:
\begin{align*}
\sum_{j=1}^m w^j &> \beta \quad (\text{significant total}) \\
w^j &< \epsilon \quad (\text{each withdrawal is small})
\end{align*}
with $\beta \in (0,1]$, $\epsilon \ll 1$.

\end{itemize}

\end{minipage}
}
\end{center}
 
\smallskip

\noindent\textbf{Formal Definition of SLID Scam.} 
\begin{defi}[SLID]
A liquidity pool is defined as a SLID scam over the time interval $[t_0, t_k]$ if the following hold:
\begin{enumerate}
    \item \textbf{Control Retention:} $\texttt{burn}(\texttt{LPT}_{\text{init}}) = \text{false}$.
    \item \textbf{Inflated Sales:} A sale sequence $\mathcal{S}$ exists such that:
    \[
    \quad q^i < \delta \cdot B^{t_i}
    \]
     
    \item \textbf{Withdrawals:} A withdrawal sequence $\mathcal{W}$ exists such that:
    \[
    \sum w^j > \beta, \quad w^j < \epsilon
    \]
    \item \textbf{Stable Appearance:} Price and volume exhibit low volatility:
    \[
    \texttt{std}(p(t)) < \theta_p, \quad \texttt{std}(V(t)) < \theta_v
    \]
    for some small thresholds $\theta_p$, $\theta_v$.
\end{enumerate}

Using the actions above, we label a pool SLID when the owner keeps control of the LP (1) and the pool still looks normal to users (4), and there is clear value extraction either through owner‑side sells (2) or owner withdrawals (3). Note that it is possible that the scammer leverages (2) and (3) in combination in an LP to exploit the pool and take profit in a SLID case.
\end{defi}
 
%SLID, as defined, exhibits slow and deceptive extraction patterns.

\smallskip
\noindent\textbf{Threat model}. A SLID adversary (i.e., LP owner) aims to:
\begin{itemize}
    \item Stealthily extract profit through long-term operations,
    \item Maintain an illusion of a healthy market (low volatility),
    \item Evade on-chain scam detectors and user suspicion.
\end{itemize}

% {\smallskip
% \noindent\textbf{Threat model}. A SLID adversary (i.e., the LP owner) may exploit LPs containing user/investor funds in DEXs by employing a gradual profit-taking strategy (combined by three core fraudulent actions) to (i) evade the current fraud detection systems; and (ii) maintain user confidence, enabling larger profits over time.}

\smallskip

\noindent{\bf Remark.} 
To ensure our SLID definition aligns with actual filtering outcomes, we conducted a second-round manual validation. From 71 SLID-like LPs, we randomly sampled 14 (20\%) and verified the presence of all three fraudulent behaviors: unburned LP tokens (action-\ding{172}), inflated token selling (action-\ding{173}), and gradual withdrawal (action-\ding{174}). All sampled LPs exhibited the full behavioral pattern, confirming the accuracy of our definition. Rather than enforcing rigid thresholds, our definition embraces the fuzzy, incomplete nature of real-world scams. Parameters such as lifespan (\textbf{P1}), profit-taking (\textbf{P2}), and total profit (\textbf{P3}) are assessed through partial or fuzzy matching, capturing variants of canonical SLID traits (see more thoughts in Appendices~\ref{appendixRationale} and~\ref{appendix:datasetreli}). This flexible design reflects the strategic behavior of rational attackers, who aim to maximize profit while delaying detection, mirroring signaling game dynamics applied in deception~\cite{casey2018deception,pawlick2021game,luca2016fake,luca2016reviews}. By emphasizing retained control, incremental drains, and price stability, our heuristic definition captures the trade-off between stealth and sustained exploitation.

%\noindent \textbf{Is our definition accurate?} To ensure that our definition aligns with the filtering designs, we conducted a second-round manual validation. From the 71 SLID-like LPs, we randomly sampled 14 instances (20\%). Each selected LP was re-examined by verifying the presence of all three fraudulent behaviors: unburned LP tokens (action-\ding{172}), inflated token selling (action-\ding{173}), and gradual withdrawal (action \ding{174}). All passed.

\smallskip

\subsection{From Definition to Rule-based Heuristic}\label{subsec:heuristic}

Following the above, we recognize that real-world scams can exhibit a wide range of behaviors (e.g., low to high profits, varying transaction frequencies, different lifetimes). To capture this variability, we designed a dynamic heuristic detection framework based on the definition and with flexible thresholds. 

% \nasrin{As part of this process, we are constructing a ground-truth dataset to validate and refine our detection criteria; this ground-truth set is broader than the SLID-like dataset obtained after filtering, as it captures a wider spectrum of behaviors indicative of SLID scams}

Our heuristic has three validators: \textit{Honeypot Validator}, \textit{Profit Validator}, and \textit{Owner Activity Validator}. Each targets different properties from the extracted data of LPs (cf. RQ1 findings). If one LP is flagged by all three validators (i.e.,  exhibiting all observed traits), it will be labelled as SLID scam.

% If one LP is flagged by all three validators (i.e.,  exhibiting all observed traits), we add it to the ground truth dataset as SLID scam.

% (i.e., broader set than SLID-like dataset after filtering) as a SLID scam.

\smallskip
\noindent\textbf{Honeypot validator.}~\label{subsubsec:honeypot} 
This component assesses whether tokens are honeypots. A successful \acro scam depends on a gradual increase in victim participation and buying volume (\textbf{P2}). 

Our \textit{honeypot validator} employs rule-based criteria derived from prior research of \cite{honeypot,honeypot2}. Instead of simply replicating exact detection methods, we apply criteria to specific features extracted from token's smart contracts (implemented in \cite{honeypotvalidator}) and set an upper threshold to identify honeypot signals.  

In particular, we extracted the following features from the paired token's smart contract: ``buy tax'', ``sell tax'', ``cannot buy'', ``cannot sell all'', ``slippage modifiable'', ``transfer pausable'', ``personal slippage modifiable'', and ``trading cooldown''. These features were evaluated based on specific criteria; each indicates different scenarios in which the LP's owner can illegally interfere with the user's activity in the LP for gaining profit.
% An upper threshold was defined to identify potential honeypots.
Here, we provide an intuitive example: an LP is flagged as a honeypot if the token's ``buy tax'' exceeds 50\%, based on tax-related criteria from earlier research~\cite{honeypot}.

\smallskip
\noindent\textbf{Profit validator.}\label{subsubsec:profitmetric}
Profit-related activities reveal \acro characteristics (\S\ref{subsubsec:proft}) from two insights: (i) the scam likelihood by evaluating realized profit, an essential element of all successful scams; (ii) the scam likelihood by evaluating the patterns of unrealized profits generated. Thus, we set both \textit{realized profit} and \textit{unrealized profit} as our metrics. Those values can be obtained through the collected DEX orders (more in \S\ref{sec:ml}). 

The LP owner's realized profit can be computed by taking the difference between the total value of their investment in the LP (in USD) and the total value that has been withdrawn over the LP's lifetime under the base token (detailed equation in Appendix~\ref{appendixeq1}). We flag an LP $\mathcal{P}$ with its owner $\mathcal{O}$ having a positive total realized profit. Although not all projects that generate realized profits are scams, our observations in \S\ref{subsubsec:proft} show that all reported scams exhibit a positive realized profit. 

The unrealized profit of the LP's owner is another critical metric, especially when the \acro LP shows a significant growth in unrealized profit after the first month of deployment. The first-month unrealized profit represents the portion of profit (in the base tokens) belonging to the owner from the LP's value after its initial month of deployment (detailed computing equation in Appendix~\ref{appendixeq2}). In \S\ref{subsubsec:proft}, we applied this unrealized profit metric for comparisons between \acro and rug pull scams, revealing a significant disparity between the average first-month unrealized profit ratios generated by the two groups (0.67\% of the initial investment for rug pull and 156\% of the initial investment for \acro LPs). 
% However, we set the threshold for the unrealized profit in this heuristic to 0 due to the uncertainty stemming from our current small initial SLID-like dataset from RQ1, with the lowest computed unrealized profit for a \acro instance being $1.3e^{-14}$.

We set the threshold for unrealized profit in this heuristic to 0, due to the limited size of our initial SLID-like dataset from RQ1. The smallest computed unrealized profit in this dataset was $1.3e^{-14}$, but given the narrow sample, we acknowledge that this value may not reflect the broader range of real-world SLID scams. While restricting unrealized profit analysis to the first month could miss later-developing cases, applying a soft heuristic threshold that requires only a positive unrealized profit during this period allows us to capture highly relevant examples that align closely with known SLID characteristics.

% We acknowledge that restricting unrealized profit analysis to the first month might miss later-developing \acro cases. Therefore, using a soft heuristic threshold of positive first-month unrealized profit ensures our ground truth dataset captures highly relevant cases aligned closely with known \acro characteristics.}

\smallskip 
\noindent\textbf{Owner activity validator.}~\label{subsubsec:profitgenmetric} 
 \acro scams prioritize a gradual approach by multiple small-proportion profit-taking actions over an extended period (\S\ref{sec:rq1}). We accordingly built \textit{Owner Activity Validator} in the following steps: 
 \begin{packeditemize}
     \item excludes LPs where owners faithfully burn their initial (LP) tokens after deploying the LP (i.e., $\texttt{burned}(\texttt{LPT}_{\text{init}}) = \text{true}$);
     \item continue to detect the LPs where their owners retain control over the initial ($\texttt{LPT}_{\text{init}}$) tokens;
     \item observe/analyse the owner's activities in exploiting profits (whether $\mathcal{O}$ applies $\mathcal{W}$ or/and $\mathcal{S}$). 
 \end{packeditemize}
 
We measure their activities through two metrics: (i) \textit{Profit-taking order count}, denoted by $c$, where $c = \texttt{count}(\mathcal{W}) + \texttt{count}(\mathcal{S})$, accounts for the total number of DEX orders involved in profit extraction; and (ii) \textit{order impact}, denoted by $I$, indicates the ratio of the $t^{th}$ profit-taking order size of value compared to the LP's total value before the activity is executed. 
%To detect LPs likely to exhibit \acro scam profit-taking patterns, 
We define their thresholds as:
\begin{equation}
 c \geq {T_\text{count}}, \, \text{where} \, I^t = \frac{q^t}{B^{t-1}} < T_\text{impact}, \, \forall \ q \  \text{by} \ \mathcal{O},
\end{equation} 
where $T_\text{count}$ and $T_\text{impact}$ are the thresholds (\S\ref{sec:rq1}) of the owner's profit-taking orders counts and maximum impact on the LP.

We set $T_\text{count}=5 $ as the lower threshold (the fewest profit-taking orders executed by $\mathcal{O}$) for flagging an LP as SLID (informed by our observations in \$\ref{sec:rq1}). Meanwhile, we define $I^t$ as $I^t = \frac{q^t}{B^{t-1}}$, where $q^t$ represents the order size of any profit-taking order made by $\mathcal{O}$ at time $t$. $B^{t-1}$ is the LP size (total base token value in the LP) right before this order.

To reflect the characteristic of small-proportion orders, we set the impact threshold of all owners' profit-taking orders to be less than 95\% ($T_\text{impact}<0.95$) of LP size. Our analysis in \S\ref{sec:rq1} revealed that 95\% was the lowest profit-taking impact recorded for a rug pull instance. It also ensures we capture a broader range of scams, including potential \acro scams that were unnoticed by rug pull detection heuristics. In comparison, for rug pull detection heuristics in \cite{rugpull,spammer}, only LPs with a 0.99 profit-taking impact are flagged.

Additionally, we decided not to incorporate adaptivity for all numerical thresholds because of the lack of ground truth. We focus on confirmed scams and mislabeled rug pulls. We acknowledge that attackers may adapt to the threshold and evade detection, but our model still lays a foundation.

\begin{takeaway1}
\noindent\textbf{Side findings}
       (\textit{How to detect SLID from LP owner activities}): An LP is flagged as a \acro scam if the owner retains control of the initial LP tokens since deployment and demonstrates a consistent pattern of profit-taking. Specifically, this includes executing 5+ profit-taking orders over the LP's lifetime, each in small proportions (with individual order impacts below 0.95).
    %\end{minipage}
%}
\end{takeaway1}
\begin{figure*}[h]
  \begin{minipage}[t]{0.33\linewidth} % Adjust the width as needed
    \centering
    \includegraphics[width=\linewidth]{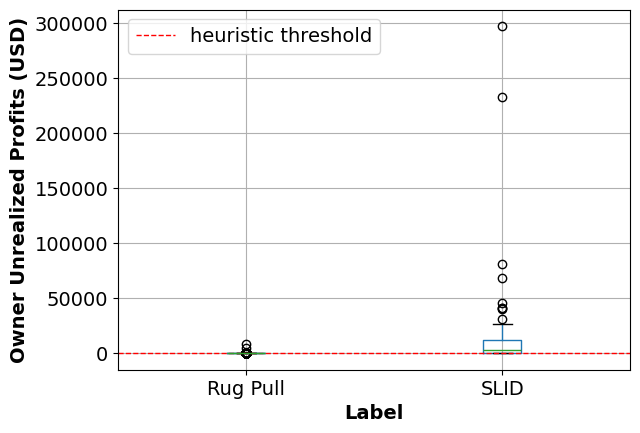}
    \caption{Distribution of owner 1st month unrealized profit.}
    \label{fig:up_compare}
  \end{minipage}%
  \hfill
  \begin{minipage}[t]{0.32\linewidth} % Adjust the width as needed
    \centering
    % You can add another figure or content here
    \includegraphics[width=\linewidth]{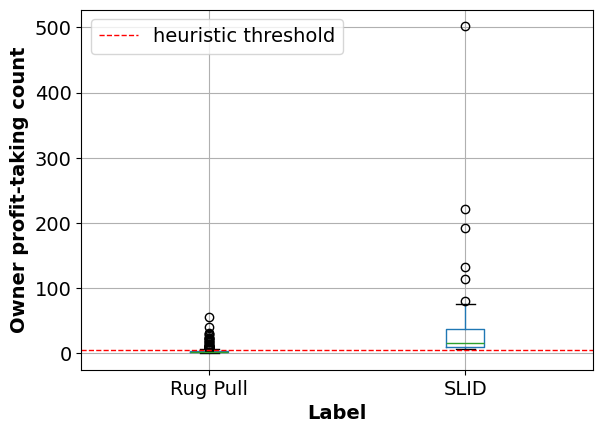} % Placeholder for the second graph
    \caption{Distribution of owner 1st month profit-taking order count.}
    %in \acro scam LPs
    \label{fig:tpcount_compare}
  \end{minipage}%
  \hfill
  \begin{minipage}[t]{0.32\linewidth} % Adjust the width as needed
    \centering
    % You can add another figure or content here
    \includegraphics[width=\linewidth]{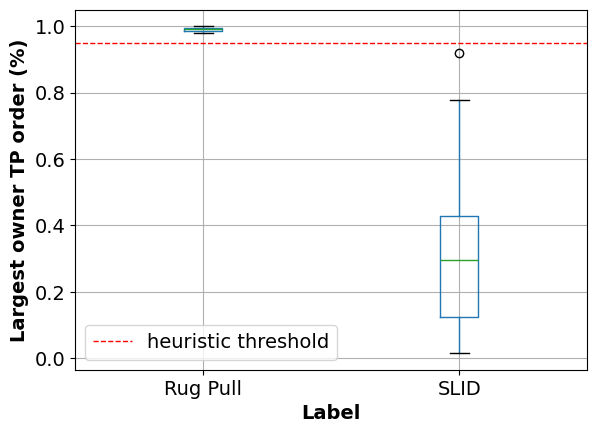} % Placeholder for the second graph
    \caption{Distribution of the largest owner profit-taking order to the LP’s size.}
    \label{fig:impact_compare}
  \end{minipage}
\end{figure*}
\smallskip

\noindent\textbf{Remark on heuristic validity.} To evaluate the reliability of the heuristic and validate the selected thresholds and cut-offs, we conducted a series of empirical assessments:

\begin{packeditemize}
    \item We first conducted the false positive rate analyses to determine how often legitimate LPs are incorrectly flagged as malicious. We applied the heuristic to 100 randomly selected legitimate LPs, each associated with Ethereum tokens whose legitimacy is strongly supported by their creation or sponsorship by large, reputable funds or companies \cite{fundnormal}. The heuristic correctly identified all non-SLID LPs as ``not SLID'', without generating any false positives. 
    
    \item We then analyse the effectiveness of the selected numeric cut-offs and thresholds in distinguishing between legitimate and illegitimate LPs. To support this, we visualize their impacts in separating LP types (Figure~\ref{fig:up_compare}--\ref{fig:impact_compare}). We use the aforementioned 100 legitimate LPs alongside illegitimate ones, including scams such as rug pulls, honeypots, and SLID, sourced from our initial DeFi scam lists (\S\ref{subsec:groundtruth1}). Beyond the initial honeypot and legitimate LP filtering, the remaining heuristics are designed to separate SLID from rug-pull LPs. These include thresholds based on:  (\textit{i}) \textbf{unrealized profits} after the first month; (\textit{ii}) \textbf{owner’s taking profit order count} after the first month; and (\textit{iii}) \textbf{impact of taking order size on the LP}. 
    \begin{packeditemize}   
        \item[$\circ$] We demonstrate that the three thresholds are effective in distinguishing SLID from rug pulls. Each figure displays a red dashed line representing the cut-off value for the respective metric. In all cases, the box-and-whisker plots for SLID LPs lie entirely on one side of the threshold, while the rug-pull LPs fall entirely on the opposite side. When combined using a logical AND condition, these three filters achieve perfect separation between SLID and rug-pull LPs in our evaluation set. No rug-pull LP satisfies all three threshold conditions simultaneously, and all SLID LPs do, resulting in 100\% accuracy for this subset.
    \end{packeditemize}
\end{packeditemize}

\begin{takeaway}
\noindent\textbf{Findings of RQ2:}
         The rule-based heuristic for \acro LP detection (with the validators for honeypot, profit, and owner actions) was demonstrated to be an effective detector, yielding 100\% accuracy against 100 legitimate LPs.
\end{takeaway}
\section{Detecting \acro on Ethereum's (RQ3)}~\label{sec:ml}
We conduct a large-scale scam detection by applying our heuristic (\S\ref{subsec:heuristic}) to datasets (including LPs, user activities) collected from six leading Ethereum DEXs (\S\ref{subsec:datacollect} and \S\ref{subsec:dexs}). We further propose an ML approach (\S\ref{subsec:ML}) to provide early warning of scams and detect borderline cases that the heuristic failed to capture (explanation in Appendix~\ref{appendixwhyml}).

\begin{figure*}[h]
  \centering
  \includegraphics[width=0.94\linewidth]{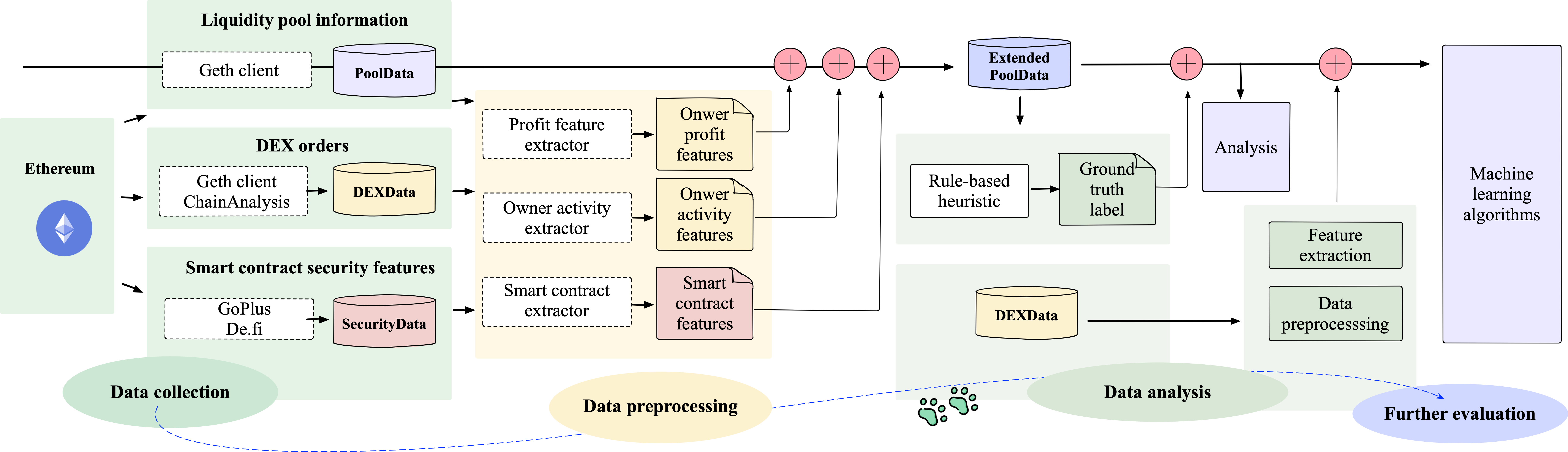}
  \caption{Overall processes of our large-scale \acro detection on Ethereum.}
  \label{fig:process}
\end{figure*}

%\nasrin{Please use this link:  https://drive.google.com/file/d/1-gDXjnesTtH5nlFfKWefCoYYgKbpBIdz/view?usp=drivesdk}

% From the designed and evaluated rule-based heuristic, we apply it to Ethereum historical data of all LPs and its activities to detect the happened \acro scams amongst the six largest DEXs on Ethereum. A machine learning method is then leveraged by the result of heuristic detection to develop an early warning system for more robust detection of \acro scam with a shorter time range required, which is vital for investor protection in such scam type with a gradual period. Figure\ref{fig:process} illustrates the process in detail that executed this section.

\subsection{Data Collection}~\label{subsec:datacollect}
We collected longitudinal Ethereum's LP data spanning over \textbf{67} months. We preprocessed the data to extract relevant features. We utilized both heuristic-based and ML algorithms to classify the LPs as \acro or non-\acro.

% We first gather all existing LPs and their DEX orders amongst the largest Ethereum DEXs and flag the LPs that triggered all three components from our rule-based heuristic. We expect this step to provide a larger ground truth dataset of \acro and data of other LPs with their respective activity information, which will be used for heuristic and machine learning detections later.

% \begin{table*}[ht]
%     \centering
%     \begin{tabular}{cc}
%         \toprule
%         \multicolumn{1}{l}{\textbf{Category}} &\textbf{Total} \\
%         \midrule
%         \multicolumn{1}{l}{\textbf{LP}} & {319,166} \\
%         \quad UniSwap & 312,299\\
%         \quad SushiSwap & 2,898\\
%         \quad Balancer & 2,819\\
%         \quad Curve & 294\\
%         \quad PancakeSwap & 443\\
%         \quad BancorSwap & 413\\
%         \midrule
%         \multicolumn{1}{l}{\textbf{DEX order}} & {917,659,580} \\
%         \quad Deposit & 12,207,714\\
%         \quad Withdraw & 5,671,266\\
%         \quad Buy & 505,364,145\\
%         \quad Sell & 394,416,456\\
%         \midrule
%         \multicolumn{1}{l}{\textbf{Smart Contract Audit}} & 302,675\\
%         \bottomrule
%     \end{tabular}
%     \caption{Summary of our collected data on different categories}
%     \label{tabstat}
% \end{table*}

\smallskip
\noindent\textbf{Liquidity pool information.}~\label{subsubsec:pool} 
We first collected all the existing LPs in Ethereum (denoted as \textsf{PoolData}) to be used for both detection methods. 
We utilized the well-known Geth client node \cite{geth} to retrieve all Ethereum block information, starting from block 6,627,000 (November 2, 2018, the date when the first LP was deployed) to block 20,081,867 (June 13, 2024, the time of writing this work). Then, we filtered only pools containing at least one well-established token (such as ETH, USDT, or USDC) as the pool's legitimate base token. We avoid adding less-recognized and less-active pairs to the dataset.  

As a result, we gathered a total of 319,166 deployed LPs across the six largest DEXs on Ethereum \cite{llama1} (Table~\ref{table:tabstat}), including Uniswap \cite{uniswap0}, Curve Finance \cite{crv0}, Balancer \cite{balancer0}, SushiSwap \cite{sushi0}, PancakeSwap \cite{pancake0}, and BancorSwap \cite{bancor0}. We extracted key features (Appendix~\ref{appendixliq}) for each harvested LP, including creation timestamps, token addresses involved in the pool, owner addresses, and the initial LP value in USD, using Chainalysis API \cite{transpose}.

% \subsubsection{LP Information}~\label{subsubsec:pool} In this step, we first collect all the existing LPs in Ethereum that will be used for both detection methods (denoted as \textsf{PoolData}). Since the first-ever LP was deployed on Nov 2, 2018, we used a well-known Geth client node to get all Ethereum block information (which is used in prior research of...) from block 6627000 (Nov 2, 2018) to block 20081867 (June 13, 2024). To enhance the collected pool's reliability, we applied a filtering condition only to take pools with at least one well-established token (such as ETH, USDT, or USDC) to become the pool's legit ``base token''. This filter avoids low-quality pairs (that contain two unknown tokens) and enhances the credibility of the dataset. From this process, we've collected 319,166 deployed LPs that match the requirements amongst the six largest DEXs on Ethereum, i.e., Uniswap, Curve Finance, Balancer, SushiSwap, PancakeSwap, BancorSwap (Ethereum exchange's Total Value Lock ranking from DefiLlama, June 2024). We extract essential LP features for each harvested LP, including creation timestamps, pool's involved token addresses, owner addresses, and initial LP value in USD using the Chainanalysis API (detailed in \ref{appendixliq}).

\smallskip
\noindent\textbf{DEX orders.}
Exchange activities and profit patterns within \textsf{PoolData} are essential for computing the profit-generating metrics (\S\ref{subsubsec:profitgenmetric}). We collected all types of exchange activities (via smart contract events) of six DEXs. Those activities include liquidity deposit, withdraw, and trade (buy/sell) between two tokens in each pool. Considering the data collection process operates independently of \textsf{PoolData} extraction, we specifically filtered the orders to include only those relevant to LPs in \textsf{PoolData}. 

We collected 917M+ DEX activities across the deployed LPs (Table~\ref{table:tabstat}). It accounts for 82.35\% of the total Ethereum trading volume historically \cite{llama2}.  The resulting dataset,  \textsf{DEXData}, contains rows where each entry represents an activity by either the LP owner or an investor. The dataset includes activities recorded up to July 13, 2024, where all pools in \textsf{PoolData} have at least one month of DEX activity history.

\smallskip
\noindent\textbf{Smart contract security features.}~\label{subsubsec:secfeatures} 
Token security-related features (such as buy/sell tax, trading lock) from the paired token's smart contracts are critical for applying the heuristic's Honeypot Validator. We utilized the API provided by GoPlus~\cite{goplus} and De.Fi \cite{dedotfi} to collect the token's security characteristics indicative of honeypot behavior, including buy/sell tax, sell restriction, and more. 

From the tokens (\textit{base} and \textit{paired}) in the collected LPs, we extracted their security features. These features were organized into a dataset, \textsf{SecurityData}. Each row represents a distinct token along with its complete set of security features.

\subsection{Large-scale \acro Detection}\label{subsec:dexs}
We described how we apply the heuristic to detect Ethereum's \acro scams. We began with data preprocessing and extracted the features for all LPs. We then conducted an in-depth analysis of the flagged \acro pools across multiple dimensions, including vitality, profit, and trend to create the first-ever SLID ground truth database.

\smallskip
\noindent\textbf{Data preprocessing.}~\label{subsubsec:pre1} 
We preprocessed the three datasets (i.e., \textsf{PoolData}, \textsf{DEXData}, \textsf{SecurityData}) to enhance the feature set and prepare for the heuristic detection phase.

For each LP instance in \textsf{PoolData}, we extracted all related DEX activities from \textsf{DEXData} to compile its complete activity history. Profit-related metrics were then computed (equations in Appendix~\ref{appendixeq2}) and added as new features to \textsf{PoolData}. These enriched metrics serve as inputs for the flagging process. Then, smart contract security features from \textsf{SecurityData} were mapped to corresponding LPs in \textsf{PoolData}.

Our preprocessing resulted in an updated dataset, called \textsf{ExtendedPoolData}, containing a comprehensive range of features (Table~\ref{tab:poolfeature}) required for heuristic detection.

% \subsubsection{Data Preprocessing} From the three collected datasets, we perform a data preprocessing step to ensure each of our collected LPs is updated with eligible features for analysis and flagging activity by the heuristic detector. For each LP instance from \textsf{PoolData} dataset, we extract all of its DEX activities from \textsf{DEXData}, resulting in the LP's all-time activity history. We then apply the proposed formulas in \S\ref{subsubsec:profitmetric} and \S\ref{subsubsec:profitgenmetric} to generate profit-related features and owner-taking-profit-related features for the LP. These calculated metrics will be updated to the \textsf{PoolData} dataset as new features for each respective LP, which will be used by the validators of the heuristic in the flagging process later. Similarly to \textsf{SecurityData} dataset, all collected security features will be grouped by tokens and passed through the honeypot validator (from \S\ref{subsubsec:honeypot}). The validator results for each token - a boolean value indicating whether the token is a honeypot - will be updated as a feature of the token's LP. From this step, each instance in \textsf{PoolData} dataset will be updated with a honeypot attribute (as boolean), which will be utilized by the heuristic in the heuristic detection phase.

\begin{table}[t]
    \centering
    \renewcommand{\arraystretch}{1}
    \caption{Collected data on Ethereum.}
    \label{table:tabstat}
    %\vspace{-0.1in}
    \begin{threeparttable}
    \resizebox{\linewidth}{!}{
    \begin{tabular}{c|c|c|c|c}
        \toprule
        \multicolumn{2}{c}{\textbf{Liquidity Pools}} &
        \multicolumn{2}{c}{\textbf{DEX orders}} &
        \multicolumn{1}{c}{\textbf{SC Features}} \\
        \cmidrule{1-5}
        \multicolumn{1}{c}{\cellcolor{gray!15} {DEXs}} & 
        \multicolumn{1}{c|}{\cellcolor{gray!15} {No. of pools}} & 
        \multicolumn{1}{c}{\cellcolor{gray!15} {Order type}} & 
        \multicolumn{1}{c|}{\cellcolor{gray!15} {No. of Txs}} & 
        \multicolumn{1}{c}{\cellcolor{gray!15} {No. of SCs}} \\
        
        \midrule
        
        UniSwap & 312,299 &  &  & \multirow{6}{*}{302,675}\\
        SushiSwap & 2,898 & Deposit & 12,207,714 &  \\
        Balancer & 2,819 & Withdraw & 5,671,266 & \\
        Curve & 294 & Buy & 505,364,145 & \\
        PancakeSwap & 443 & Sell & 394,416,456 & \\
        BancorSwap & 413 & & & \\
        \midrule
        \multicolumn{1}{c}{Total} & \multicolumn{1}{c}{\textbf{319,166}} & \multicolumn{1}{c}{} & \multicolumn{1}{c}{\textbf{917,659,580}} & \textbf{302,675} \\
        
        \bottomrule
    \end{tabular}
    }
      \begin{tablenotes}
       \footnotesize
       \item[] \text{Abbreviation}: \textbf{Tx}: transaction; \textbf{SC}: smart contract; \textbf{No.}: number. 
     \end{tablenotes}
  \end{threeparttable}
  \vspace{-0.2in}
\end{table}

\smallskip
\noindent\textbf{\acro detection for Ethereum.}
We applied the proposed rule-based heuristic to all instances in the \textsf{ExtendedPoolData} dataset. The new dataset now contains all required features for processing with the heuristic's validators. 

\begin{packeditemize}
\item (\textbf{Quick access to our dataset}). Out of our 319,166 selected LPs, we flagged 3,117 instances of \acro scams (accessed in \cite{slidpoolsdata}). We further conducted various steps to evaluate the reliability of the dataset (deferred to Appendix~\ref{appendix:datasetreli}):
   
\end{packeditemize}
% \subsubsection{Happened \acro on Ethereum} After the data preprocessing step, we applied our rule-based heuristic to all collected LPs collected in \textsf{PoolData} dataset - which now has all required features for processing with the heuristic's validators. From 319,166 selected LPs, we have flagged 3,117 happened \acro scams. We randomly selected 100 \acro pools for evaluation with manual observation, which showed no sign of false positives. We further verified the argument that this scam tactic is new and has not been captured previously by testing those randomly chosen pools with existing heuristics of honeypot and rug pull, and the mentioned heuristics detected none of them.

\begin{figure*}[h]
  \begin{minipage}[t]{0.29\linewidth} % Adjust the width as needed
    \centering
    \includegraphics[width=\linewidth]{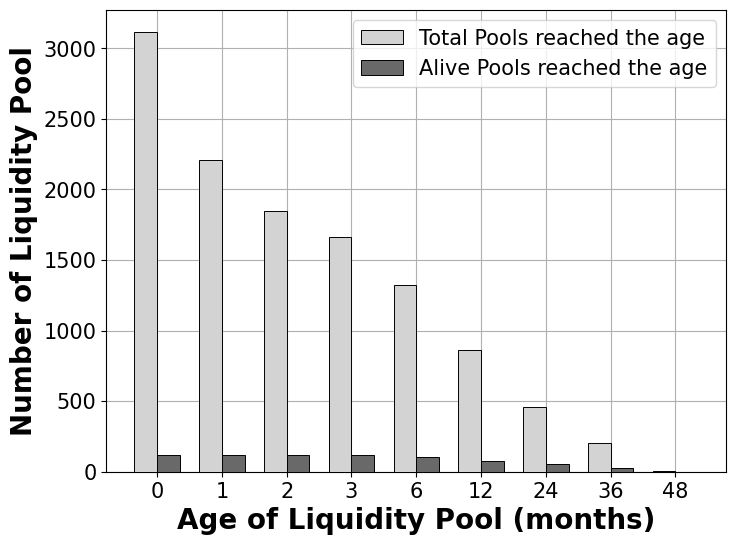}
    \caption{Lifetime analysis of captured \acro LPs.}
    \label{fig:age}
  \end{minipage}%
  \hfill
  \begin{minipage}[t]{0.33\linewidth} % Adjust the width as needed
    \centering
    % You can add another figure or content here
    \includegraphics[width=\linewidth]{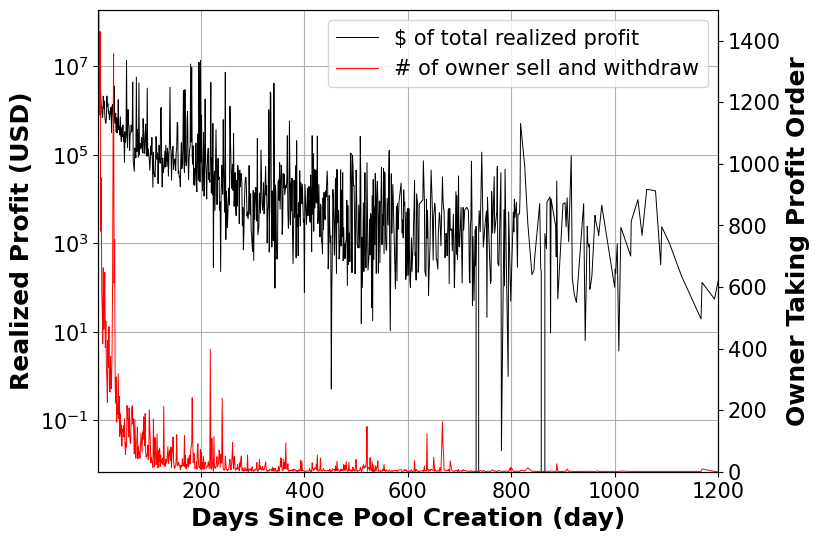} % Placeholder for the second graph
    \caption{Daily realized profit drained and scammer's profit-taking activity count.}
    %in \acro scam LPs
    \label{fig:profitscammer}
  \end{minipage}%
  \hfill
  \begin{minipage}[t]{0.33\linewidth} % Adjust the width as needed
    \centering
    % You can add another figure or content here
    \includegraphics[width=\linewidth]{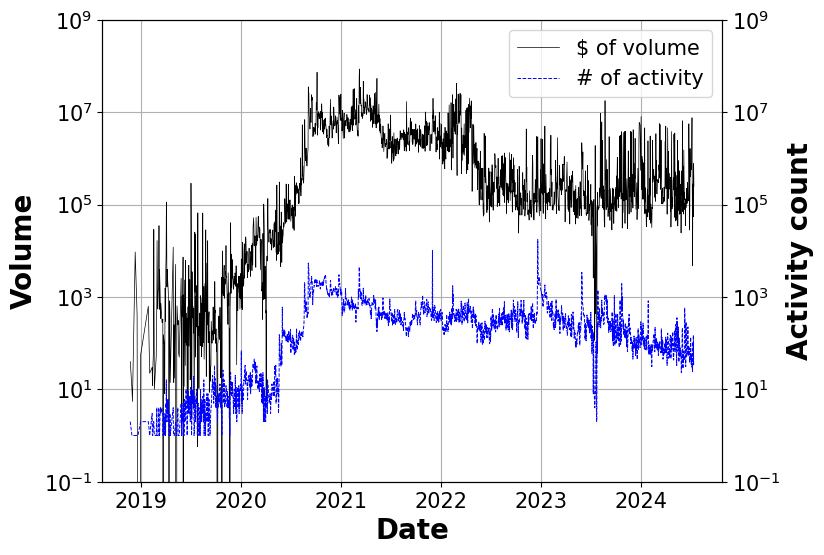} % Placeholder for the second graph
    \caption{Daily volume (in USD) and activity count in \acro scam LPs.}
    \label{fig:volact}
  \end{minipage}
\end{figure*}

\smallskip
\noindent\textbf{(\textit{i}) Vitality analysis of \acro pools.} Figure~\ref{fig:age} depicts the \acro LP age and the remaining active LP numbers in each age group. 
Of the 3,117 \acro LPs, 70.8\% (2,207/3,117) remained active after the first month, and 53.3\% (1,660/3,117) were still operational three months after their deployment. Five pools reached the longest recorded age of four years. Among the 196  pools that were active for over three years, 28 are still alive today.  The performance is exceptional compared to other scams (rug pull and honeypot).  Over their lifetime, \acro pools attracted a total of 1.49M activities on DEXs, with a cumulative volume of \$4.49B. 

We concluded that \acro scams exhibit the characteristics of a long LP lifespan, with extended periods of activity and deep user engagement (further supported by trading volume).

% \textbf{Vitality Analysis of Happened \acro pools.} The happened \acro scams exhibit its gradual nature with a long LP lifetime overall. Fig \ref{fig:age} has demonstrated the overall and alive (up to present - July 2024) LP age statistic. Over 3,117 flagged LPs, 70.8\% (2207/3,117) remained active after the first month, and 53.3\% (1660/3,117) were still alive after the first three months from their deployed day. Notably, five \acro pools reach the longest recorded age of four. These statistics are outstanding compared to the average liveliness of other scams, such as rug pull (ref) and honeypot (ref), ... Among 196 instances that have been active for more than three years, 28 of them are still alive up to the present. In the \acro pools lifetime, they attracted 1.49M DEX activities on DEXs with a total volume reached \$4.49B. This analysis highlights the \acro longevity with extraordinarily long active period and huge user involvement demonstrated by the total engaged volume.\\

\smallskip
\noindent\textbf{(\textit{ii}) Profit analysis of \acro pools.} We analyzed the financial earnings of scammers. Across all flagged \acro LPs, the scammers have lured 167,632 victims, generating \$79.1M in profit through a gradual profit-taking strategy. Additionally, unrealized profit assets worth \$24.1M remain under the control of scammers in active pools. On average, scammers achieved a profit of 6.08 times their initial investment. The top LPs generating the highest profit are described in Table~\ref{tab:prof}. The 3,117 \acro scams were initiated by 2,951 distinct addresses, with 90 addresses involved in more than one scam and one address deploying a peak of 23 \acro scams (Table~\ref{tab:scammer1}). 

We further investigated the scammers' gradual strategy by analyzing their profit-taking patterns from the first day of LP deployment. Figure~\ref{fig:profitscammer} illustrates the total number of profit-taking actions and the corresponding realized profit by days from each pool's respective deployment date. We observed that most profit-taking actions occur in the first 100 days of the \acro LPs, with 35.71\% (31,912/89,364) of sell or withdraw actions executed during this period. While the number of profit-taking actions decreases, the total realized profit remains relatively stable throughout the evaluation period, ranging between \$1K and \$100K, with a few outliers.

We concluded that the scammers, who leveraged the SLID strategy, have the alarming ability to continue generating profits even during later stages with fewer profit-taking actions.

%\smallskip
\noindent\textbf{(\textit{iii}) Trend analysis of \acro pools.} We analysed the trend of collected pools. Figure~\ref{tab:slid_lps} illustrates a steady increase in the number of \acro pools created annually, with the exception of 2024, where the lower number is due to the limited capture period (6 months). Figure~\ref{fig:volact} illustrates the total activities and volume associated with \acro scams. We describe the development of \acro scams below.
 \begin{packeditemize}
  \item \textbf{\textit{Early development phase (2018-2020):}} During this period, a small number of \acro pools were created, exhibiting low volume and minimal activity. This could be due to the limited presence of DEXs and the lack of user familiarity with decentralized activities \cite{dexempi}. However, there was a gradual increase in the development of \acro LPs, with recorded increments in user engagement.
    % \item \textit{\textbf{Early Development Phase (2018 - 2020):}} This period experienced a small number of \acro scam pools created and low volume and activities committed. This may be due to the lack of DEXs and fewer users' familiarity with decentralized activities (cite). However, there was a gradual development of \acro scam LPs with recorded increments in engagement.
    \item \textbf{\textit{Exploding phase (Mid-2020-2023):}} Following the rapid growth of DeFi, the market experienced a significant surge in \acro scams. The number of \acro LPs created in 2020 was over 10 times greater than in 2019, accompanied by an explosion in trading volume and activity within these pools, with volumes ranging between
     \$10K and \$100M. 
    This phase also saw record-breaking performances for \acro scams, with an all-time peak total volume of \$85.7M on March 8, 2021, and peak DEX activity counts of 17,586 on December 20, 2022, across all \acro LPs.
    \item \textbf{\textit{Stabilized phase (2023-present):}} The statistics for \acro LPs maintained high levels of stability, with daily activity counts ranging from 100 to 1000 and daily volumes ranging from \$10K to \$10M. During this period, the average daily volume was \$608,159, and there were 266.67 DEX activities per day across \acro pools. Although not exhibiting the same rapid growth as the previous one, the scam tactic remains attractive to scammers, as evidenced by the creation of 1,107 new pools. This continued appeal is likely due to the lack of prior recognition.
\end{packeditemize}

 \begin{takeaway1}
    \noindent We establish the first-ever ground truth dataset of SLID LPs, consisting of 3,117 (out of 319,166) pools. The dataset serves as the foundation for ML analysis. We also observe a clear upward trend in the initialization of SLID scams over the years, suggesting that this strategy is being increasingly adopted by malicious actors.
  \end{takeaway1}
% We conclude that there is a clear upward trend in the initialization of \acro scams over the years. The SLID strategy has been increasingly adopted by malicious actors.

% From this section, the first-ever SLID LPs ground truth is obtained and analyzed, then will be used in the ML section below.

\subsection{ML Approach for Early \acro Warning}\label{subsec:ML}
We take a step further. Our heuristic flags \acro pools only \textbf{after} profit-taking actions that trigger the heuristic's validators. However, investors have already lost their money. 

We thus propose a \underline{proactive}, ML-based approach to raise early warnings and prevent users from engaging during the pool's \textbf{early stages}. In addition, the ML model is also effective in identifying \textbf{borderline SLID cases} that may be missed by rule-based heuristics (detailed explanation in Appendix~\ref{appendixwhyml}). In particular, we extracted features from the LP in progressively shorter time windows to identify a reasonable timeframe that offers high prediction accuracy.

\begin{comment}

\begin{table}[!htp]
    \centering
    \caption{Statistic of the created \acro LPs over years.}
    \label{createdyears}
    \resizebox{\linewidth}{!}{
    \begin{tabular}{c|ccccccccc}
       \toprule
        \multicolumn{1}{c}{\textbf{Year}} & 
        \rotatebox{70}{\text{2017}} &
        \rotatebox{70}{\text{2018}} &
        \rotatebox{70}{\text{2019}} &
        \rotatebox{70}{\text{2020}} & 
        \rotatebox{70}{\text{2021}} & 
        \rotatebox{70}{\text{2022}} & 
        \rotatebox{70}{\text{2023}} & 
        \rotatebox{70}{\text{\makecell{2024\\(-June)}}} \\
        \midrule
        \textbf{\acro LPs} &  \cellcolor{gray!15}  {23} & \cellcolor{gray!15}  {41} & \cellcolor{gray!15} {52} & \cellcolor{gray!15} {585} & \cellcolor{gray!15} {446} & \cellcolor{gray!15} {863} & \cellcolor{gray!15} {979} & \cellcolor{gray!15} {128} \\
        \bottomrule
    \end{tabular}
    }
\end{table}

\end{comment}

\begin{figure}[!]
\centering
\resizebox{\linewidth}{!}{
\begin{tikzpicture}
\begin{axis}[
    width=13cm, % Plot width
    height=5cm, % Plot height
    ybar,
    bar width=0.35cm, % Adjust bar width
    symbolic x coords={2017, 2018, 2019, 2020, 2021, 2022, 2023, 2024},
    xtick=data,
    xticklabel style={rotate=30, anchor=east, yshift=-2pt},
    xlabel={Year (\textit{p.s.} 2024 up to June)},
    ylabel={SLID LPs},
    ymin=0, ymax=1200,
    nodes near coords,
    every node near coord/.append style={yshift=5pt},
    nodes near coords align={center},
    enlarge x limits=0.2, % Adjust spacing between bars
    legend style={at={(0.5,1.25)}, anchor=north, legend columns=-1}
    background color=gray!60, % Set the background color
    grid=both, % Optional: Add grid lines for better readability
    grid style={dotted} % Style for the grid lines
]

% Bar with solid color (Early Development Phase)
\addplot[pattern=north east lines, pattern color=blue] coordinates {
    (2018, 64)
    (2019, 52)
    (2020, 585)
    (2021, 446)
    (2022, 863)
    (2023, 979)
    (2024, 128)
};
% Legend entries
%\legend{( liquidy pools in different phases )}
\end{axis}
\end{tikzpicture}
}
\vspace{-0.1in}
\caption{Statistic of the created SLID LPs over years.}
\label{tab:slid_lps}
\vspace{-0.2in}
\end{figure}

\smallskip
\noindent\textbf{Data preprocessing and feature extraction.}~\label{subsubsec:pre2}
We labelled all LPs and utilised the labelled data for training ML algorithms. The dataset contains 3,117 flagged \acro pools and 316,049 non-\acro pools in the \textsf{ExtendedPoolData} dataset.

For early detection, we constrained the liquidity data (i.e., DEX activities) used for feature extraction to the first \textit{$d$} days after the deployment of pools in \textsf{ExtendedPoolData}. During the preprocessing phase, we extracted features for each LP within the first \textit{$d$} days of its creation and appended these features to the corresponding LP records in \textsf{ExtendedPoolData}.

We then utilized the resulting dataset, \textsf{$d^{th}$-day}, to train and evaluate ML algorithms. We varied the \textit{$d$} value, training the algorithms separately for each duration. We determine the shortest time frame by systematically reducing \textit{$d$}, which balances high accuracy in detecting \acro scams with early warning capabilities. For pools with a lifetime shorter than \textit{$d$}, features were extracted based on all available DEX activities from deployment to their inactive state.

To build an effective classifier within the specified timeframe, we extracted 57 features from \textsf{DEXData}, capturing patterns of DEX activities (Appendix~\ref{appendixfeatures}). These features were categorized as: \textit{Owner Activity Features (OAF)}, \textit{User Activity Features (UAF)}, \textit{LP Features (LPF)}, and \textit{Profit Features (PF)}.

\begin{figure*}[htb]
    \centering
  \subfigure[\textbf{Accuracy}]{
        \resizebox{0.45\linewidth}{!}{
        \begin{tikzpicture}
            \begin{axis}[
               width=11cm, % Plot width
               height=4cm, % Plot height
                ybar,
                bar width=0.12cm,
                symbolic x coords={267, 150, 100, 60, 59, 58, 57, 56},
                xtick=data,
              %  nodes near coords,
              %  nodes near coords align={vertical},
                ymin=0.85, ymax=1.01,
                x tick label style={rotate=45,anchor=east},
                legend style={at={(0.5,1.50)},
                anchor=north,legend columns=-1},
            ]
            \addplot coordinates {(267, 1.00) (150, 1.00) (100, 1.00) (60, 1.00) (59, 1.00) (58, 1.00) (57, 1.00) (56, 1.00)};
            \addplot coordinates {(267, 0.96) (150, 0.96) (100, 0.96) (60, 0.96) (59, 0.95) (58, 0.95) (57, 0.95) (56, 0.93)};
            \addplot coordinates {(267, 0.94) (150, 0.94) (100, 0.94) (60, 0.94) (59, 0.94) (58, 0.93) (57, 0.94) (56, 0.93)};
            \addplot coordinates {(267, 0.94) (150, 0.94) (100, 0.94) (60, 0.94) (59, 0.94) (58, 0.93) (57, 0.94) (56, 0.93)};
         %   \legend{Heuristic, Random Forest, Logistic Regression, XGBoost}
            \end{axis}
        \end{tikzpicture}
        }
    }
\vspace{-1pt} 
    \subfigure[\textbf{Precision}]{
    \resizebox{0.45\linewidth}{!}{
        \begin{tikzpicture}
            \begin{axis}[
                width=11cm, % Plot width
               height=4cm, % Plot height
                ybar,
                bar width=0.12cm,
                symbolic x coords={267, 150, 100, 60, 59, 58, 57, 56},
                xtick=data,
              %  nodes near coords,
              %  nodes near coords align={vertical},
                ymin=0.85, ymax=1.01,
                x tick label style={rotate=45,anchor=east},
                legend style={at={(0.5,-0.20)},
                anchor=north,legend columns=-1},
            ]
            \addplot coordinates {(267, 1.00) (150, 1.00) (100, 1.00) (60, 1.00) (59, 1.00) (58, 1.00) (57, 1.00) (56, 1.00)};
            \addplot coordinates {(267, 0.96) (150, 0.95) (100, 0.94) (60, 0.94) (59, 0.96) (58, 0.95) (57, 0.95) (56, 0.94)};
            \addplot coordinates {(267, 0.94) (150, 0.94) (100, 0.95) (60, 0.94) (59, 0.95) (58, 0.93) (57, 0.92) (56, 0.92)};
            \addplot coordinates {(267, 0.95) (150, 0.95) (100, 0.94) (60, 0.96) (59, 0.95) (58, 0.94) (57, 0.94) (56, 0.92)};
          %  \legend{Heuristic, Random Forest}
            \end{axis}
        \end{tikzpicture}
        }
    }
\vspace{-1pt} 
    \subfigure[\textbf{Recall}]{
        \resizebox{0.45\linewidth}{!}{
        \begin{tikzpicture}
            \begin{axis}[
               width=11cm, % Plot width
               height=4cm, % Plot height
                ybar,
                bar width=0.12cm,
                symbolic x coords={267, 150, 100, 60, 59, 58, 57, 56},
                xtick=data,
              %  nodes near coords,
              %  nodes near coords align={vertical},
                ymin=0.85, ymax=1.01,
                x tick label style={rotate=45,anchor=east},
                legend style={at={(0.5,-0.20)},
                anchor=north,legend columns=-1},
            ]
            \addplot
            coordinates {(267, 0.95) (150, 0.89) (100, 0.85) (60, 0.78) (59, 0.77) (58, 0.77) (57, 0.77) (56, 0.76)};
            \addplot 
            coordinates {(267, 0.97) (150, 0.96) (100, 0.96) (60, 0.97) (59, 0.97) (58, 0.96) (57, 0.96) (56, 0.92)};
            \addplot 
            coordinates {(267, 0.96) (150, 0.95) (100, 0.94) (60, 0.95) (59, 0.96) (58, 0.94) (57, 0.93) (56, 0.94)};
            \addplot 
            coordinates {(267, 0.95) (150, 0.96) (100, 0.93) (60, 0.94) (59, 0.93) (58, 0.94) (57, 0.96) (56, 0.93)};
         %   \legend{Logistic Regression, XGBoost}
            \end{axis}
        \end{tikzpicture}
    }
    }
\vspace{-1pt} 
    \subfigure[\textbf{F1 score}]{
        \resizebox{0.45\linewidth}{!}{
        \begin{tikzpicture}
            \begin{axis}[
               width=11cm, % Plot width
               height=4cm, % Plot height
                ybar,
                bar width=0.12cm,
                symbolic x coords={267, 150, 100, 60, 59, 58, 57, 56},
                xtick=data,
              %  nodes near coords,
              %  nodes near coords align={vertical},
                ymin=0.85, ymax=1.01,
                x tick label style={rotate=45,anchor=east},
                legend style={at={(0.5,-0.20)},
                anchor=north,legend columns=-1},
            ]
            \addplot
            coordinates {(267, 0.97) (150, 0.94) (100, 0.92) (60, 0.88) (59, 0.87) (58, 0.87) (57, 0.87) (56, 0.86)};
            \addplot 
            coordinates {(267, 0.96) (150, 0.95) (100, 0.95) (60, 0.95) (59, 0.96) (58, 0.95) (57, 0.95) (56, 0.93)};
            \addplot 
            coordinates {(267, 0.95) (150, 0.94) (100, 0.94) (60, 0.94) (59, 0.95) (58, 0.93) (57, 0.92) (56, 0.93)};
            \addplot 
            coordinates {(267, 0.95) (150, 0.95) (100, 0.93) (60, 0.95) (59, 0.94) (58, 0.94) (57, 0.95) (56, 0.92)};
            \end{axis}
        \end{tikzpicture}
    }
    }
    \caption{\textbf{Detection evaluation metrics over varying days}. Notations: X axis for \textit{days}, Y axis for \textit{values}; \textit{Heuristic} in blue (first bar), \textit{Random Forest} in red (second), \textit{Logistic Regression} in brown (third), and \textit{XGBoost} in black (fourth).}
    \label{fig:detection-metrics}
\end{figure*}

\smallskip
\noindent\textbf{ML algorithms.} We framed the early detection problem as a binary classification task, focusing on models that deliver high accuracy while maintaining low complexity. Specifically, the model selection criteria emphasized efficiency, model complexity, computational cost, and accuracy, requiring a careful balance among these factors. 

Based on these considerations, we focused on three ML models: \textit{Random Forest} \cite{randomf}, \textit{XGBoost} \cite{xgb}, and \textit{Logistic Regression} \cite{log}.  We exclude several alternative approaches, such as graph-based or network-based strategies, due to their high complexity and computational demands \cite{rfbetter, graphweakness}. Moreover, they are unsuitable for the scenarios we address~\cite{graphweakness} (more rationales in Appendix~\ref{appendixRationale}).

\smallskip
\noindent\textbf{Early detection.} 
We measured their effectiveness across various \textit{$d$} values using metrics of accuracy, precision, and recall. 

To address the imbalance problem, we applied class weighting \cite{scikitweight} and utilized GridSearchCV from \texttt{sckikit-learn}~\cite{scikit} for hyperparameter tuning for each $d$-algorithm combination.  To demonstrate the detection efficiency in a shorter time period, we conducted an experiment using the heuristic on the 3,117 pools (pseudocode provided in Algorithm~\ref{pseudo:ml}). 

We selected the initial value \textit{$d$} as 267 days because, within this time frame, all four detectors remain optimal.

\smallskip
\noindent\textbf{Detection results} (Figure~\ref{fig:detection-metrics}). 
As the time range was reduced, the performance of recall of the heuristic began to decline due to the limited number of order activities, which caused the heuristic validators to fail to trigger some detections. Given that the true negative count in the ground truth (316,049) is substantial compared to the true positive (3,117), both accuracy and precision of the heuristic remained unchanged when \textit{$d$} decreased. In contrast, the three ML algorithms maintained stable and good performances with decreasing time ranges. However, Random Forest performs consistently better than the other two ML algorithms in most \textit{$d$}.

Random Forest maintained an accuracy greater than 93\% down to \textit{$d$} = 56, whereas Logistic Regression and XGBoost saw their accuracy drop to 93\% when using data within a time range of 58 days from the LP's deployment date. Starting from \textit{$d$} = 56, the accuracy of all ML algorithms fell to 93\%, marking the threshold at which all algorithms lost their high performance observed previously.

Overall, the three ML algorithms demonstrated detection capabilities that were 4.77 times faster than the proposed heuristic, achieving high accuracy with just 57 days of data compared to the heuristic's requirement of 268 days.

\begin{takeaway}
\noindent\textbf{Findings (RQ3):}
       Our rule-based heuristic for Ethereum DEXs identifies 3,117 \acro pools. Our ML-enhanced approach, particularly Random Forest, improves 4.77× performance, reaching 95\% accuracy in just 57 days.
\end{takeaway}

\section{Discussion}

%We acknowledge several limitations in future works.

 % \noindent\textbf{\textit{\textcolor{red}{$\bullet$} Time-dependent Heuristic.}} The rule-based heuristic operates effectively with a complete history of DEX activities, specifically those exceeding 280 days of data. While the ML algorithms have addressed this limitation by reducing the required historical data to 57 days, this dependency makes it impractical to develop a real-time detector capable of identifying \acro scams at an earlier stage.  \nasrin{I am not sure if we need to have this as a limitation}
 
 % \noindent\textbf{\textit{\textcolor{red}{$\bullet$} Data Constraints.}} \textsf{PoolData} and \textsf{OrderData} datasets have the limitation of not capturing the complete history of all pools and DEX activities on Ethereum, as we focused on collecting data from only the six largest DEXs on this network. However, the selected DEXs account for 82.35\% of the total DEX volume across the network, indicating that data from the non-selected exchanges is unlikely to influence the study's results significantly. \nasrin{I am not sure if we need to have this as a limitation}

\subsection{\acro Mitigation}
SLID remains concealed and challenging for users to detect quickly. However, we stress several key points worth noting.

\begin{packeditemize}
     \item \textbf{Verify token purpose and cross-check the claims.} Ensure the token aligns with its intended purpose (utility, governance, investment). Any discrepancies between official claims and token behavior may signal a scam.
     \item \textbf{Evaluate LP tokens.}
         Confirm that initial LP tokens are burned or securely locked using trustless mechanisms. Tools like Etherscan, DEXTools, and Chainalysis can help with verification. Unburned/unlocked tokens are red flags.
     \item \textbf{Assess on-chain data.} Prioritize transparency by thoroughly evaluating on-chain activities. Regularly monitor the tokens owned by LP owners for any signs of gradual drain. 
\end{packeditemize}

\subsection{Comparison to ``Similar'' Concepts}

\noindent\textbf{Conventional stock dilution.} SLID scams closely resemble unauthorized stock issuance in traditional finance, where inflating the supply of shares without shareholder approval is a clear form of fraud. Regulatory bodies like U.S. Securities and Exchange Commission (SEC) enforce rules, which requires shareholder consent before issuing additional stock \cite{nysedefinition, morganstanleyequity}. A notable case is the SEC’s 2016 enforcement action against Goldman Sachs for facilitating naked short selling~\cite{nysedefinition}. Similarly, in SLID, token issuers retain unilateral control over liquidity or token supply, diluting existing holdings without transparency or user consent - practices that would be considered illegal in traditional markets \cite{seccase1,seccase2}.

\smallskip
\noindent\textbf{Arbitrage bot.} Those bots are automated agents designed to exploit price discrepancies across different markets or pools, often executing rapid trades to capture profit without taking directional market risk. Their behavior may involve frequent interactions with LPs, such as buying low in one and selling high in another, but they operate within transparent, rule-based systems and do not manipulate token supply or LP ownership. Unlike SLID scammers, arbitrage bots do not retain special privileges over token minting or LP tokens, and their profits stem from legitimate market inefficiencies rather than exploitative control. Therefore, despite their aggressive trading patterns, arbitrage bots are fundamentally different from SLID scams in both mechanism and intent.

\smallskip
\noindent\textbf{Memecoins.} We use the recently popular Trump token as a representative example of memecoins. While it shares superficial traits with SLID scams, such as rapid insider accumulation, it does not meet our criteria for SLID. In this case, insiders and affiliates acquired their tokens directly from the LP shortly after its launch (approximately two minutes post-deployment). Importantly, these tokens were not freely minted or allocated without cost. Instead, they were purchased at the prevailing market price, meaning that insiders assumed genuine financial risk and did not benefit from disproportionate or zero-cost supply advantages \cite{trumpcaseclarify}.

\smallskip
% In a simple way, we can easily identify their differences from a financial view.
Their differences are evident from a financial perspective.

%\begin{center}
%\fbox{%
%\begin{minipage}{0.9\linewidth}
%\begin{packeditemize}
\begin{itemize}
    \item \textbf{Asset origin:} 
    SLID scammers exploit the system by either minting tokens without disclosure or retaining their initial LP tokens, enabling them to later extract profits fraudulently \textbf{at no cost}. In contrast, legitimate traders acquire assets transparently at market prices and cannot mint unlimited tokens or withdraw the entire pool's value.
    \item \textbf{Risk exposure:} Legitimate traders bear genuine market risk, whereas SLID scammers face minimal risk due to the \textbf{control over} token minting and huge liquidity.
    \end{itemize}

\subsection{Adversarial Adaptation and Real-world Integration}

\noindent{\textbf{Advarsarial adaptaion and trade-offs.}} While our detection methods are effective against canonical SLID patterns, adaptive adversaries may attempt to evade detection by modifying their extraction strategy. One plausible approach is to slow down profit-taking activities or reduce the number of withdrawal or sell actions below our heuristic thresholds (e.g., fewer than five profit-taking orders or extended delay beyond 60 days). Although this behavior may temporarily evade our rule-based validator, it introduces a critical trade-off: the scammer must significantly reduce their short-term return on investment (ROI) and prolong their attack window, increasing the risk of manual discovery or external scrutiny. In contrast, our machine learning model, trained on early-stage LP activity, remains effective even when profit-taking is delayed. This is because it captures structural anomalies such as retained LP token control, disproportionate minting of paired tokens, lack of LP token burning, and abnormal trading volume patterns. Thus, delaying extraction may reduce signal strength marginally but does not meaningfully bypass ML detection within the first 57 days.

Another adaptive strategy involves distributing fraudulent actions across multiple EOAs or using a different address from the deployer, thereby masking the scammer’s identity and mimicking legitimate user behavior. While this tactic may degrade the performance of purely wallet-based heuristics, the aggregate behavioral patterns (e.g., centralized token control, orchestrated sell timing, and highly correlated transaction profiles) remain detectable by our ML classifier. We emphasize that both heuristic and ML models exhibit adversarial robustness by design: the former requires multiple observable actions to trigger, while the latter generalizes across behavioral patterns rather than specific identities. Nonetheless, we recognize the value of formal adversarial resilience testing. Future work could quantify the trade-off between scam profitability and detection delay. This analysis would further highlight the dilemma faced by SLID scammers: maximizing stealth often requires sacrificing profitability or operational efficiency.

\smallskip
\noindent{\textbf{Real-world integration.}} Our solution can integrate directly with Ethereum DEXs as an LP monitoring and early warning service, providing real-time alerts on suspicious liquidity pools and enabling proactive fraud detection. By extracting key features from token transaction histories, it continuously tracks evolving on-chain behaviors while scaling efficiently to handle diverse market conditions. In addition, integration with payment service providers allows our system to evaluate wallet credibility by identifying addresses linked to SLID creation and analyzing suspicious transaction histories, even when legitimate tokens like Bitcoin are involved. This combined approach enhances risk assessment, blocks malicious token flows before they affect end users, and supports the development of stronger DeFi security standards.

\section{Conclusion}

This paper introduces slow liquidity drain (SLID) scams as a stealthy and financially devastating threat to the DeFi ecosystem. By combining heuristic-based detection with machine learning, we empirically identified 3,117 SLID pools, exposing \$103.2M in cumulative losses. Our analysis focuses on Ethereum-based DEXs and the six largest platforms, which together account for 82.35\% of total DEX volume. While this excludes smaller DEXs and other blockchains with different trading dynamics, the core SLID behaviors we identify are representative of dominant market activity and provide a strong foundation for future cross-chain analysis. We hope this study raises community awareness to prevent such daily losses. ``Slow'' does not mean safe; it silently drains your wallet.

\section*{Ethics Considerations}

Our research focuses on uncovering and mitigating the slow liquidity drain (SLID) scam within the DeFi ecosystem. Given the sensitive nature of this study, we adhered to established ethical guidelines to ensure responsible data handling, minimize harm, and promote fairness across all stakeholders.

%In this work, we leverage approximately 917M DEX transactions/activities on the 5 largest DEXs on Ethereum. We focused our research on the behavior pattern of the owner that creates \acro liquidity pool and their activity. All examined data are publicly available by the Ethereum network, with all addresses anonymized by only keeping the first/last 5 characters of each, and generated mapping values for similarities determination (of determining transactions involving the same tokens/pools/users) in analysis tasks. We have no intention of deanonymizing any of the involved addresses for privacy reasons. 

\smallskip
\noindent\textbf{Privacy considerations.}
We focused our research on the behavior pattern of the owner that creates \acro liquidity pool and their activities. All examined data are publicly available on the Ethereum network, with all addresses anonymized by only keeping the first/last 5 characters of each, and generated mapping values for similarity determination (of determining transactions involving the same tokens/pools/users) in analysis tasks. We have no intention of deanonymizing any of the involved addresses for privacy reasons.

\smallskip
\noindent\textbf{Respect for people.}
We prioritized user privacy throughout this research. All data analyzed was sourced from publicly available blockchain records, and no personally identifiable information (PII) was collected or processed. While blockchain data is inherently pseudonymous (also claimed before), we ensured that our findings respected user anonymity. Additionally, any insights derived from community-reported rug pulls were aggregated to avoid targeting individual users.

\smallskip
\noindent\textbf{Beneficence.}
To maximize benefits while minimizing potential harm, we aimed to alert the DeFi community to the risks posed by SLID scams. Our research introduces early detection mechanisms, equipping users with tools to make informed decisions and reduce financial losses. 

To mitigate the risk of misuse, we avoided publishing details that could enable malicious actors to refine or replicate SLID scams. Instead, the study emphasizes preventive strategies and raises awareness among users and researchers.

\smallskip
\noindent\textbf{Justice.}
The benefits of this research are distributed equitably across the DeFi community. By identifying a previously underexplored scam type, we empower both users and platform operators to take proactive measures. Our findings are not intended to single out specific entities but rather to foster collective improvements in security and trust.

\smallskip
\noindent\textbf{Respect for law and public interest.}
This study was conducted in compliance with relevant regulations surrounding data privacy and blockchain technology. All activities were limited to publicly accessible data, and no vulnerabilities were exploited during the research. Instead, we aim to enhance user protection through transparency and collaboration with the broader DeFi community.

\smallskip
\noindent\textbf{Proactive community engagement.}
Throughout this study, we engaged with community stakeholders, including affected users and DEX platforms, to validate our findings and refine detection strategies. By fostering open dialogue, we sought to align our objectives with the broader goal of improving DeFi security while addressing community concerns.

By adhering to these ethical principles, our research aims to strike a balance between innovation and responsibility, contributing to a safer and more resilient DeFi ecosystem.

\bibliographystyle{unsrt}
\bibliography{bib}

\appendix

\section{Appendix}

\subsection{Additional Related Work}
\label{sec:rw}

\noindent\textbf{Liquidity pool, DEX and DeFi.}
Over the past five years, the rise of DeFi and DEXs has attracted millions of users and billions in inflows, leading to numerous research studies \cite{defiopp,defitail,tradesurvey,dexempi,dexamm,univ31,univ32,arbitrageloss}. Hendrik et al. \cite{defiopp} examined the opportunities and challenges for DeFi and its products in this mass adoption era. Fan et al. \cite{tradesurvey} conducted a comprehensive study on cryptocurrency trading activities for measuring market valuation. Jianlei et al. \cite{dexempi} investigated the DEXs roles and their impact on the development of the DeFi ecosystem.
Saleh et al. \cite{univ31} analyzed the effects of liquidity on Uniswap for liquidity providers and conducted estimations of fees and impermanent loss. Similarly, Lioba et al. \cite{univ32} developed a theoretical model to analyze risks and returns faced by liquidity providers under various circumstances.

\smallskip
\noindent\textbf{Cryptocurrency scam detection.}
Various research efforts have focused on analyzing DEX scams, such as rug pull \cite{rugpull,spammer,lin2024crpwarner,tradeortrick} and honeypot \cite{honeypot,honeypot2,honeypot4}. For instance, Bruno et al. \cite{rugpull} utilized ML algorithms to identify rug pull and scam-likely tokens. Cernera et al. \cite{spammer} conducted a longitudinal analysis of the lifetime of rug pull LPs on Ethereum and Binance Smart Chain. Rundong et al. \cite{honeypot2} introduced various types of honeypot traps on DEXs along with their respective detection methods. Similarly, 
 Christof et al. \cite{honeypot} conducted a comprehensive honeypot analysis with a proposed symbolic-execution-based solution. In addition, the scams originating from traditional finance, such as pump-and-dump schemes \cite{pd1,pd2} and Ponzi schemes \cite{sadponzi,ponzi2}, have also been studied to provide warnings to investors. Several studies  \cite{graph1,graph2,graph3,graph4} have suggested general fraud detection models that identify anomalous transactions or users through graph neural networks. However, among all the studies in this field, the \acro scam has yet to be discovered.

%======================================
\subsection{LP Owner Financial Guarantee (cf. \$\ref{sec:bck})}
\label{appendixproof}
%======================================

Let the LP's initial deposit (by the owner) be $x$ units of the paired token and $y$ units of the base token. From the formula of the AMM model from Uniswap \cite{ammformula}, the relationship between these tokens in the LP is reflected by:

\begin{equation}
    x \times y = k
\end{equation}
where $k$ is a constant determined by the initial liquidity and remains unchanged. With the mentioned setup, the pool's owner initially monopolizes the paired token supply, which means no paired tokens $x$ exist outside the LP before the pool's deployment. Any investor/user that is investing/interacting with the LP must first buy the paired token from the LP by swapping it with base tokens. Let $\Delta x$ represent the total amount of the paired token bought by users or distributed as incentives (for users that deposit a pair of tokens for liquidity providers). The formula will be rewritten to represent the pool's state at any moment after deployment:

\begin{equation}
\begin{split}
    (x + \Delta x) \times (y + \Delta y) = k\\
    x' \times y' = k
\end{split}
\end{equation}
where $\Delta y$ is the change of base token amount in the LP due to the change of paired token, and x'/y' is the amount of paired/base token in the LP after the change, compared to the initial deposit amount (from the owner). Under the AMM model, the constant $k$ must not change.

In the worst scenario where no investor interacts with the LP ($\Delta x = 0$) or all investors sell back their paired token ($= \Delta x$, since the monopoly of paired tokens from the owner made the limit of paired token that investor possess totally at any time cannot be larger than $\Delta x$) to the pool, the base token possessed by the owner afterward is computed by:

\begin{equation}
    y' = \frac{k}{x'} = \frac{x \times y}{x + \Delta x - \Delta x} = \frac{x}{x} \times y = y
\end{equation}

This proves that in the worst-case scenario, where the LP has no investor or all investors that previously bought (and used the bought paired token to deposit for incentives) sell back all their received paired tokens to the pool, the base token balance controlled by the owner will still be equal to $y$. Hence, with this setup, the LP owner is guaranteed to maintain their original investment of base tokens $y$, which contain real value, under any sequence of investor interactions after deployment.

\subsection{Computing Equation}\label{appendixeq1}

\noindent\textbf{Realized profit computing equation.} To compute the realized profit, we let $m$, $n$, $o$, $p$ represent the total number of owners' sell, withdraw, buy and deposit orders to their pool, respectively. The realized profit can be defined as follows:
\begin{equation}
    \begin{split}
        \textit{Profit\_realized} = & \left( \sum_{i=1}^{m} Sell^i + \sum_{j=1}^{n} Withdraw^j \right) \\
        & - \left( \sum_{k=1}^{o} Buy^k + \sum_{l=1}^{p} Deposit^l + \text{GAS} \right)
    \end{split}
\end{equation}
where the owner's investment activity (which adds assets in base tokes to the pool) includes $Buy^k$ and $Deposit^l$, representing the asset value (in USD) added to their LP during the $k^{\text{th}}$ buy and $l^{\text{th}}$ deposit orders, respectively. In contrast, the owner's realized return consists of $Sell^i$ and $Withdraw^j$, which are the asset values in base token (in USD) that the owner removed from the LP during the $i^{\text{th}}$ sell and $j^{\text{th}}$ withdraw orders, respectively. The metric also accounts for the total network gas fee (\text{GAS}) used for LP deployment and all DEX activities performed by the owner. Since the equation focused on computing realized profit, all of the variables are only around the owner's balance change of the base token (which holds the real value) through their activity in the LP (which also involves paired tokens). 

\smallskip
\noindent\textbf{Unrealized profit computing equation.}\label{appendixeq2}
To compute this unrealized profit, we begin with the pool's value (in USD), denoted as $x$, in the base token, and $owner\_share$, representing the LP owner's portion of the pool's asset. During the LP deployment phase, $x$ will be the owner's initial deposit value: $x = x_0 = Deposit_0$ and $owner\_share$ is 1, as they initially own 100\% of the pool. For each subsequent $t^{th}$ DEX order made to the LP, the value of $x$ is updated as follows:

\begin{equation}\label{eq:poolbalance}
    x = x^t = x_{0} + \sum_{p=1}^{t} y^p
\end{equation}
where $y^p$ represents the value of \textit{base tokens} (USD) for the $p^{th}$ DEX order. If the order is a buy or deposit (indicating a positive $y$), it signifies that $y$ USD worth of base token has been added to the LP through the order. Conversely, sell and withdraw orders indicate a decrease in value (indicating a negative $y$) due to the outflow of base tokens. Further, for each ``deposit'' and ``withdraw'' order, the LP proportions shared amongst the liquidity providers change by,

\begin{equation}\label{eq:owner-share}
    owner\_share^t = 
    \begin{cases}
        \begin{array}{l}
            owner\_share^{t-1} \times \frac{x^{t-1}}{x^t} + \frac{y^t}{x^t} \\
            \text{(if $t^{\text{th}}$ is a deposit/withdraw activity}\\
            \text{from the owner),}
        \end{array} \\
        \begin{array}{l}
            owner\_share^{t-1} \times \frac{x^{t-1}}{x^t}\\
            \text{(if $t^{\text{th}}$ is a deposit/withdraw activity}\\
            \text{from other investors).}\\
        \end{array}
    \end{cases}
\end{equation}

Eq.~\ref{eq:owner-share} indicates that ${owner\_share}^t$ increases if the owner deposits more into the pool, and conversely, it decreases if the owner executes withdraw orders. Similarly, the owner's share will decrease if a user deposits into the LP and will increase if a user withdraws from it. The mechanism behind the owner pool's share in Eq.~\ref{eq:owner-share} is explained in more detail in \S\ref{appendixeqex}. Therefore, the LP's owner's first-month unrealized return is defined as follows:
\begin{equation}
    \begin{split}
         \textit{Profit\_unrealized}_{1m} = \textit{Return\_unrealized}_{1m}\\
         = {x}^t \times {owner\_share}^t,\\
         \text{where $\textit{Profit\_realized}_{1m} \geq 0$}
    \end{split}
\end{equation} 
where $t$ in this equation represents the order number executed in the LP on the DEX since its deployment to the end of the first month. If the LP has a positive realized profit threshold (having reached the break-even point and already generated profits), we can consider the unrealized return as unrealized profit since it represents a profit beyond the realized gains.

%========================================
\subsection{Owner's LP Share}
\label{appendixeqex}
%========================================

We explain how the owner's liquidity share is updated after a deposit or withdrawal in a DEX LP. We denote the share of owners at the $t^{th}$ order as $owner\_share^t$. We present two scenarios.

\begin{packeditemize}
\item 
\textbf{Deposit/withdraw excluding the owner.}\label{subsec:explainfor}
When a deposit or withdrawal order is executed in the LP and the sender is not the owner, the owner's share is updated as follows:

\begin{equation*}
    owner\_share^t = owner\_share^{t-1}\times\frac{x^{t-1}}{x^t}.
\end{equation*} 

The owner's LP share before the $t^{th}$ order is calculated by multiplying their previous share $owner\_share^{t-1}$ by the LP value before the order $x^{t-1}$. After the $t^{th}$ order, the updated share is determined by dividing the owner's liquidity amount by the new LP value $x^t$.

\item 

\smallskip
\textbf{Deposit/withdraw by the owner.}
% When the LP owner execute the order $t^{th}$ with size $y_t$, we compute the new owner share as follows:
When the LP owner executes the $t^{th}$ order with size $y^t$, we compute the new owner share as follows:
\begin{equation*}
    owner\_share^t= owner\_share^{t-1}\times\frac{x^{t-1}}{x^t} + \frac{y^t}{x^t}.
\end{equation*} 
% Whether $t^{th}$ order is a deposit (positive $y^{th}$) or withdraw (negative $y^{th}$), $y_t/x^t$ is the formula to calculate the percentage of this new liquidity part added to the LP value that belongs to the owner, when combined with the owner share before $t^{th}$ (explained in \ref{subsec:explainfor}) and its newly updated share from $t^{th}$ order, the new owner share can be calculated. If the $t^{th}$ order is the deposit activity from the owner, the combined part $\frac{y_t}{x^t}$ will also have a positive value (since LP value ${x^t}$ is always a positive value) and makes the owner share increase after the updated, and vice-versa for withdraw activities from the owner.

Whether the $t^{th}$ order involves a deposit or withdrawal, the ratio $\frac{y^t}{x^t}$ represents the percentage of the new liquidity part contributed by the owner relative to the pool's value. This ratio, reflecting either an addition or subtraction, merges with the owner's share prior to the $t^{th}$ order and updates their stake accordingly. If the $t^{th}$ order is a deposit by the owner, the ratio $\frac{y^t}{x^t}$ yields a positive value, as the LP value $x^t$ is inherently positive, thereby increasing the owner’s share post-update. Conversely, in the case of withdrawals, this ratio becomes negative, diminishing the owner’s share.

\end{packeditemize}

\subsection{Dataset Reliability Evaluation  (cf. \$\ref{sec:ml})}\label{appendix:datasetreli}

We also evaluated the reliability of our collected ground truth that was labeled from the data collection process.
\begin{itemize}
    \item (manual observation) We selected 100 \acro pools randomly and our analysis revealed no sign of false positives. For every pool we reviewed, we first verified it with well‑established scam traits: deployers ($i$) still held the full proportion of LP tokens; ($ii$) executed a sequence of profit-taking sequence that stepped the price down; and ($iii$) had never burned liquidity. Next, we cross‑checked these pools against our declarative filters to guarantee all heuristic signals are detected in each flagged LPs. Our manual review found zero false positives where all 100 pools satisfied scam traits and heuristic conditions. 
    \item (external cross-check) We tested the 100 randomly selected SLID pools using existing honeypot and rug pull heuristics, none of which detected the scam. Moreover, we randomly collected 20 verified rug‑pull tokens from ChainAbuse \cite{chainabuse} and 20 verified honeypots from GoPlus \cite{goplus} and cross-matched their contract addresses with our 3,117 SLID labels. The intersection was empty (both false positive/negative $=0$), indicating perfect precision and recall ($=1$) and showing that our filter does not misclassify other scam types as SLID.
\end{itemize}

%======================================
\subsection{The Role of ML in SLID Detection (cf. \S\ref{subsec:ML})}\label{appendixRationale}
%======================================

\subsubsection{Why ML is essential for our SLID detection}\label{appendixwhyml}
We have two main reasons as below.

\noindent\textbf{($i$) Early detection when SLID pools fail to form required characteristics.} In the rule-based heuristic, certain features indicative of SLID scams take time to form, as the scam must exhibit multiple defining characteristics before it can be detected through our fuzzy heuristic. These features, such as gradual liquidity withdrawal, sales of inflated tokens, or lifetime, are often subtle at first and accumulate over time. As a result, heuristic methods may struggle to identify SLID scams in their early stages.

Machine learning offers an advantage by detecting early-stage SLID cases that have not yet fully formed with required characteristics. ML models are capable of recognizing patterns in data that may seem insignificant to human-defined rules, such as early signs of unusual behavior, even when those behaviors are not yet obvious. The early detection capability was validated in our experiments, where the ML model's prediction time was 4.77 times faster than the heuristic approach (Figure~\ref{fig:detection-metrics}).

\smallskip

\noindent\textbf{($ii$) Detecting borderline cases.} ML can be used for detecting borderline cases or characteristics by adjusting its threshold. In this study, we define a borderline SLID pool as an LP that satisfies only a subset of the formal SLID conditions (Definition 1, \S\ref{sec:definition}), such as:

\begin{packeditemize}
    \item The pool's owner is taking profit but less than 5 times ($1 \leq T_{count}<5 $).
    \item The pool yielded zero profit in the first month, but became positive and continues to grow in subsequent periods.
\end{packeditemize}

Since this paper sets out the first evidence‑based foundation for SLID detection, every filter in the heuristic of Definition 1 (\S\ref{sec:definition}) is served with a fixed range of thresholds (based on the analysis from \S\ref{sec:rq1}) to minimize false positives. As a result, such borderline pools continue to pose risks to liquidity providers, given that their subtle or noisier behavioral patterns tend to evade detection by rigid rule-based methods.

Unlike rule-based heuristics, which rely on rigid thresholds and predefined patterns (although we've relaxed the thresholds for more real-time capturing enhancement), ML approaches can implicitly incorporate uncertainty and behavioral noise \cite{mlexplain1,mlexplain2}. They excel at identifying patterns in data that may be too complex or nuanced. By leveraging training on large, diverse datasets, ML models can detect these early, emerging signs of SLID scams even before the fraudulent activity becomes fully apparent or fits within a fixed heuristic definition. Multiple previous studies on DeFi scams have successfully employed machine learning techniques to detect borderline cases \cite{nftrug, ponzi2, rugpull}.

\subsubsection{Rationale for selected MLs}
We chose \textit{Random Forest}, \textit{XGBoost}, and \textit{Logistic Regression} because they offer:

\begin{packeditemize}
    \item \textbf{efficiency for real-time detection}: Those models process transactions quickly and are computationally efficient once trained, ideal for (near-)real-time scenarios. They also scale well to high-throughput cases (e.g., memLP monitoring, transactional analysis), where rapid predictions on incoming data are critical.
    
    \item \textbf{compatibility with feature-based data}: They excel in structured, tabular data environments, leveraging numerical and categorical features without requiring complex data transformations, unlike graph-based approaches (which require transforming data into relational structures).
    
    \item \textbf{balance of complexity and accuracy}: Random Forest and XGBoost capture non-linear patterns. Logistic Regression provides a lightweight yet robust alternative for simpler cases, with high performance and manageable complexity.
    
    \item \textbf{transparency and interpretability}: These models provide clear insights into feature importance (Random Forest) or coefficients (Logistic Regression), aiding decision-making and enabling user trust in the detection process.
\end{packeditemize}

%======================================
\subsection{Additional Statistics (cf. \S\ref{subsec:dexs})}%\acro LPs with Most Profit
\label{appendixprofit}
%======================================

Table~\ref{tab:prof} presents the top 5 LPs with the highest profits. 
Table~\ref{tab:scammer1} shows the scammers with the most \acro LPs.

\begin{table}[http]
    \centering
    \renewcommand{\arraystretch}{1}
    \caption{LPs with the highest total profit generated}
    \label{tab:prof}
    \resizebox{\linewidth}{!}{
    \begin{tabular}{>{\centering\arraybackslash}p{3.5cm} |  >{\centering\arraybackslash}p{3.5cm} c}
        \toprule
       \multicolumn{1}{c}{\textbf{LP address}} & \multicolumn{1}{c}{\textbf{Paired Token}} &\textbf{Profit (USD)}\\
        \midrule
       0x32Ce...cb81b &  CORE & \$4,261,805\\
       0xBE5...e0732 & OptionRoom & \$3,673,710\\
       0x092...DB485 &  Public Mint &\$2,470,875\\
       0x8d9...c7f5F & Finxflo & \$1,174,583\\
       0x9dc...198Fd & Bondly Token & \$986,684\\
        \bottomrule
    \end{tabular}
    }
\end{table}

\begin{table}[http]
    \centering
    \renewcommand{\arraystretch}{1}
    \caption{\acro scammers that deployed most \acro LPs}
    \label{tab:scammer1}
    \resizebox{\linewidth}{!}{
    \begin{tabular}{>{\centering\arraybackslash}p{3.5cm} |>{\centering\arraybackslash}p{3.5cm}  c}
        \toprule
        \multicolumn{1}{c}{\textbf{Scammer address}} &\textbf{\acro scam created}&\textbf{Profit (USD)}\\
        \midrule        
        0x3DC...1f887&23&\$17,307,795\\
        0x4bb...ca0a0&15&\$395,054\\
        0x285...409c1&9&\$298,459\\
        0x521...dCDD2&7&\$2,064\\
        0x06c...AD87C&6&\$4,170\\
        \bottomrule
    \end{tabular}
    }
\end{table}

%======================================
\subsection{Data Collection and Features (cf. \S\ref{subsec:datacollect})}
%======================================

\noindent\textbf{Collected data.} We report the statistics of collected data (Table~\ref{table:tabstat}) used in the data collection process (cf. \S\ref{subsec:datacollect}).

\smallskip
\noindent\textbf{Gathered features.}\label{appendixliq}
We show features (Table~\ref{tab:poolfeature}'s \textit{upper}-side) from collected LPs and DEX order events from \textsf{PoolData}. We gather token security features for honeypot determination.

\smallskip
\noindent\textbf{Extracted features in ML detectors.}\label{appendixfeatures}
We show 57 features (Table~\ref{tab:poolfeature}'s \textit{bottom}-side) extracted from \textsf{DEXData} within a given time window $d$ for the training and testing activities of the ML detector. The features are categorized as:

\begin{packeditemize}
    \item \textit{Owner activities features} (\textbf{OAF}). These features reflect the owner's interaction behavior with their LPs, including their activity statistics within the time range of \textit{$d$}.
    \item \textit{User activities features} (\textbf{UAF}). This category includes metrics that capture user activities during this period, providing information about both the users and their activities.
    \item \textit{LP features} (\textbf{LPF}). These illustrate the characteristics of LPs during the specified period, encompassing aspects of the pool's lifetime, volume, and asset value.
    \item \textit{Profit features} (\textbf{PF}). They indicate the profit patterns generated by LP for its owner, covering a wide range of metrics, including return, profit, and return on investment.
\end{packeditemize}

%\subsection{Collected Features from DEX Order Event}\label{appendix1}
%Table \ref{tab:dexfeatures} presents the features collected from each DEX activity record in \textsf{PoolData} dataset. These features are utilized for metrics calculation in \ref{subsubsec:pre1} for heuristic application and feature extraction in \ref{subsubsec:pre2} in ML detector training.

%\subsection{Collected Security Features of Token for Honeypot Determination}\label{appendix2}
%Table~\ref{tab:sc} describes the features of the token smart contracts collected in Sec.\ref{subsubsec:secfeatures}. These features are then used to determine whether the token is associated with honeypot activities.

%======================================
\subsection{Pseudocode}
\label{appendixpseudo}
%======================================

Our main program consists of three functions (Algorithm~\ref{pseudo:lp_analys}), each focusing on an aspect of the LPs' operations.

%**************************ML**************************

\begin{algorithm}
\caption{ML approach for early SLID warning}\label{algo-ml}
\small 
\SetAlgoLined
\KwData{\texttt{DEXData}}

\SetKwProg{Fn}{\textcolor{blue}{ML\_approach }(\texttt{SLID\_pools, DEXData})}{:}{}
\Fn{}{
    \ForEach{\texttt{d} in 1 to 300}{
        \texttt{ExtendedPoolData\_[d]} $\gets$ \texttt{Extract features from DEXData in transaction range between day 1 to day d} \\ 
        \texttt{results[d][heuristic]} $\gets$ \texttt{evaluation metrics (accuracy, precision, recall) when applying heuristic on ExtendedPoolData\_[d]}\\
        \textcolor{olive}{/* get all SLID pools */}\\
        \ForEach{\texttt{model} in [RandomForest, LogRegression, XGBoost]}{
            \texttt{trained\_model} $\gets$ \texttt{Train model using ExtendedPoolData\_[d]}\\
            \texttt{evaluation} $\gets$ \texttt{Evaluate trained\_model on validation/test set}\\
            \texttt{results[d][model]} $\gets$ \texttt{evaluation metrics (accuracy, precision, recall)}\\
            
        }
    }
}
\label{pseudo:ml}
\end{algorithm}

The first function (\textsf{LP\_Age\_Analysis}) determines the longevity and activity status of each pool. It retrieves transactions associated with each LP and calculates the pool's age by comparing the earliest and latest transaction timestamps. This data helps assess how long each LP has been operational. The function also tracks whether pools are still active by checking for recent transactions.

%***********************Vitality Analysis***************************

\begin{algorithm}
\caption{Liquidity pool analysis}\label{pseudo:lp_analys}
\small 
\SetAlgoLined

\KwData{\texttt{ExtendedPoolData}, \texttt{DEXData}}

\BlankLine

\texttt{SLID\_pools} $\gets$ \texttt{ExtendedPoolData.label == 'SLID'}  
\\   \textcolor{olive}{/* get all SLID pools */}\\

Token\_Pool\_Age\_Analysis (\texttt{SLID\_pools, DEXData}) 
%\textcolor{olive}{// details in \ref{pseudo:lp_analys_detail}} \\

LP\_Profit\_Analysis (\texttt{SLID\_pools, DEXData}) 
%\textcolor{olive}{// details in \ref{pseudo:lp_analys_detail}} \\

LP\_Trend\_Analysis (\texttt{SLID\_pools, DEXData}) 
%\textcolor{olive}{// details in \ref{pseudo:lp_analys_detail}} \\

\BlankLine

\SetKwProg{Fn}{\textcolor{blue}{LP\_Age\_Analysis} (\texttt{SLID\_pools, DEXData})}{:}{}
\Fn{}{
    \ForEach{\texttt{pool} in \texttt{SLID\_pools}}{
        \texttt{pool\_txs} $\gets$ \texttt{all entries in DEXData where DEXData.pool\_address == SLID\_pools.pool\_address}\\ 
        \textcolor{olive}{/* get all transaction of the LP */}\\
        \texttt{pool\_age} $\gets$ $max(\texttt{pool\_txs.timestamp}) - min(\texttt{pool\_txs.timestamp})$\\ 
        \textcolor{olive}{/* LP age calculation */}\\
        \texttt{ages[pool\_age]} $\gets$ $\texttt{ages[pool\_age]} + 1$ \\
        \textcolor{olive}{/* add to list for graphing */}\\
        \texttt{alive[pool\_age]} $\gets$ $\texttt{ages[pool\_age]} + 1$ if max(\texttt{pool\_txs.timestamp}) $\geq$ now \\
         \textcolor{olive}{/* record alive pools of age groups */}
    }
}   
\SetAlgoLined

\SetKwProg{Fn}{\textcolor{blue}{LP\_Profit\_Analysis }(\texttt{SLID\_pools, DEXData})}{:}{}
\Fn{}{
    \ForEach{\texttt{pool} in \texttt{SLID\_pools}}{
        \texttt{pool\_txs} $\gets$ \texttt{all entries in DEXData where DEXData.pool\_address == SLID\_pools.pool\_address \& DEXData.sender == SLID\_pools.pool\_owner \& DEXData.category == sell/withdraw \& DEXData.token\_address == SLID\_pools.base\_address}\\
        \textcolor{olive}{/* get pool's owner profit\_taking activity */}\\
        
        \ForEach{\texttt{tx} in \texttt{pool\_txs}}{
            \texttt{volume[transaction.timestamp]} $\gets$ $tx[amount\_token] \times tx[token\_price]$\\
            
            \texttt{count[transaction.timestamp]} $\gets$ $\texttt{count[transaction.timestamp]} + 1$\\
        }
    }
}
\SetAlgoLined
\SetKwProg{Fn}{\textcolor{blue}{LP\_Trend\_Analysis} (\texttt{SLID\_pools, DEXData})}{:}{}
\Fn{}{
    \ForEach{\texttt{pool} in \texttt{SLID\_pools}}{
        \texttt{pool\_txs} $\gets$ \texttt{all entries in DEXData where DEXData.pool\_address == SLID\_pools.pool\_address \& DEXData.sender != SLID\_pools.pool\_owner \& DEXData.token\_address == SLID\_pools.base\_address}\\
        \textcolor{olive}{/* get all user activity of the LP */}\\
        
        \ForEach{\texttt{tx} in \texttt{pool\_txs}}{
            \texttt{volume[transaction.timestamp]} $\gets$ $tx[amount\_token] \times tx[token\_price]$\\
            
            \texttt{count[transaction.timestamp]} $\gets$ $\texttt{count[transaction.timestamp]} + 1$\\
        }
    }
}
\label{pseudo:lp_analys_detail}
\end{algorithm}

In the second function (\textsf{LP\_Profit\_Analysis}), the focus shifts to the financial activities of the LP owners. This part of the algorithm targets transactions where the LP owners are selling or withdrawing assets. By analyzing these transactions, the function calculates the total volume of assets moved, providing insights into the profit-taking behaviors of LP owners.

The final function (\textsf{LP\_Trend\_Analysis}) examines the engagement of other users within the pools, excluding the owners. It analyses transactions made by users to understand the general activity level and trends within each pool. This function helps to highlight user participation and can indicate the popularity of the LP based on user activity.

We further present pseudocode for our enhanced method by using machine learning (Algorithm~\ref{algo-ml}). 

Over 300 days, our algorithm extracts daily transaction features to evaluate and compare heuristic and machine learning models (RandomForest, Logistic Regression, XGBoost) based on accuracy, precision, and recall.

\begin{table*}[t]
    \centering
    \caption{Summerized features for datasets.}
    \label{tab:poolfeature}
    \renewcommand{\arraystretch}{1}
    \vspace{0.05in}
    \resizebox{\textwidth}{!}{
    \begin{tabular}{l|cl|cl}
        \toprule
        
         \multicolumn{1}{c}{} & \multicolumn{1}{c}{\textbf{Feature}} & \multicolumn{1}{c}{Description} &  \multicolumn{1}{c}{\textbf{Feature}} & \multicolumn{1}{c}{Description}\\
         
        \midrule
        
        \multicolumn{5}{c}{\cellcolor{gray!15} Features collected for \textbf{LPs}, \textbf{DEX activitivies} and \textbf{honeypot deterniniation}}\\
        
        \midrule
        
        \multirow{4}{*}{\rotatebox{90}{\textbf{LP}}}
        & $Address_{pool}$&Ethereum address of LPs & $Address_{base}$&Ethereum address of base token in LPs\\
        & $Address_{paired}$&Ethereum address of paired token in LPs & $Address_{owner}$&Ethereum address of the pool owner \\
        & $CreatedTime_{token}$&Created time of the paired token & $CreatedTime_{pool}$ & Deployed time of LP\\
        & $DEX$&DEX where the liquidity pool deployed & $Name$&Paired token's name\\

        \midrule

        \multirow{7}{*}{\rotatebox{90}{\textbf{DEX activities}}} 
        & $Block$ & Block number that the activity occurred & $Timestamp$&Timestamp that the activity occured \\
        & $Hash$&Transaction hash of the activity & $Category$& Activities (buy/sell/deposit/withdraw) \\
        & $Address_{pool}$&Ethereum address of LP & $Address_{sender}$&Ethereum address of the sender \\
        & $Address_{base}$&Ethereum address of base token in LPs & $Address_{paired}$&Ethereum address of paired token in LPs \\
        & $x_{paired}$ & Pool balance of paired token after activities & $y_{paired}$&Token amount of paired token\\
        & $x_{base}$&Pool balance of base token after activities & $y_{base}$& Token amount of base token  \\
        & $Price_{paired}$&Price of paired token recorded in the block  & $Price_{base}$&Price of base token recorded in the block  \\

        \midrule
        
        \multirow{5}{*}{\rotatebox{90}{\textbf{\makecell{Honeypot}}}} 
        & $Tax_{buy}$&Buy tax of the token & $Tax_{modify}$&  Owner's capability to change buy/sell \\
        & $Tax_{sell}$&Sell tax of the token & $Buyable$& User's capability to buy tokens\\
        & $SellAll$& User's capability to sell all tokens & $BalanceChange$& Owner's capablity to change user's balance \\
        & $TradingCooldown$& Time limit for tokens between activities
        & $TradingPausable$&  Owner's capbility of pausing trading\\
        & $AntiWhale$&   token activities with high amounts\\
       
        \midrule

%  \multicolumn{1}{c}{} & \multicolumn{1}{c}{\textbf{Feature}} & \multicolumn{1}{c}{Description} &  \multicolumn{1}{c}{\textbf{Feature}} & \multicolumn{1}{c}{Description}\\

     \multicolumn{5}{c}{\cellcolor{gray!15} Extracted features of \textbf{OAF}, \textbf{UAF}, \textbf{PF} and \textbf{LPF} for ML detector} \\
   
    \midrule
    
    \multirow{2}{*}{\rotatebox{90}{\textbf{OAF}}} 
    & $Owner_{dep}$  & Total owner deposit activity    & $Owner_{buy}$  & Total owner buy activity \\
    & $Owner_{with}$  & Total owner withdraw activity & $Owner_{sell}$  & Total owner sell activity \\
    
    \midrule
    
    \multirow{8}{*}{\rotatebox{90}{\textbf{UAF}}} 
    & $User_{dep}$  & Total users deposit activity & $User_{with}$  & Total users withdraw activity \\
    & $User_{buy}$  & Total users buy activity & $User_{sell}$  & Total users sell activity \\
    & $User_{count}$  & Total pools user count \\
    & $User_{countfirst}$ & User count first day of pool deployment  & $User_{counthigh}$  & User count of day with most pool users\\
    & $User_{countlast}$  & User count last day of pool deployment  & $User_{countlow}$  & User count of day with least pool users \\
    
    & $RUser_{firstonhigh}$  & The ratio of $User_{countfirst}$ and $User_{counthigh}$ & $RUser_{lastonlow}$  & The ratio of $User_{countlast}$ and $User_{countlow}$ \\
    & $RUser_{firstonlow}$  & The ratio of $User_{countfirst}$ and $User_{countlow}$ & $RUser_{firstonlast}$  & The ratio of $User_{countfirst}$ and $User_{countlast}$ \\
    & $RUser_{lastonhigh}$  & The ratio of $User_{countlast}$ and $User_{counthigh}$ & $RUser_{lowonhigh}$  & The ratio of $User_{countlow}$ and $User_{counthigh}$ \\
    
    \midrule
    
    \multirow{8}{*}{\rotatebox{90}{\textbf{PF}}} 
    & $Owner_{i}$  & Total owner investment (USD)  & $Owner_{u}$  & Total owner unrealized return (USD)  \\
    & $Owner_{r}$  & Total owner realized return (USD) & $Owner_{total}$  & Owner pool total profit/loss (USD) \\
    & $ROwner_{toni}$  & The ratio of $Owner_{total}$ and $Owner_{i}$  & $ROwner_{roni}$ & The ratio of $Owner_{r}$ and $Owner_{i}$ \\
    & $ROwner_{roi}$  & The ratio of ($Owner_{r}$$-$$Owner_{i}$) and $Owner_{i}$  &  $ROwner_{uonr}$ &  The ratio of $Owner_{u}$ and $Owner_{r}$  \\
    & $ROwner_{uoni}$  & The ratio of $Owner_{u}$ and $Owner_{i}$  \\
    & $Impact_{min}$  & Lowest pool impact of single owner's action & $RImpact_{minonavg}$  & The ratio of $Impact_{min}$ and $Impact_{avg}$\\
    & $Impact_{max}$  & Highest pool impact of single owner's action & $RImpact_{maxonavg}$  & The ratio of $Impact_{max}$ and $Impact_{avg}$\\
    & $Impact_{avg}$  & Average pool impact of all owners' actions & $RImpact_{minonmax}$  & The ratio of $Impact_{min}$ and $Impact_{max}$\\

    \midrule
    
    \multirow{12}{*}{\rotatebox{90}{\textbf{LPF}}} 
    & $Age$  & LP age ($=d$ if the pool still active) & $IsAlive$  & Whether the pool is still active\\
    & $Vol_{f}$  & First day DEX volume   & $Pval_{f}$  & LP's total value after the first day   \\
    & $Vol_{l}$  & Last day DEX volume & $Pval_{l}$  & LP's total value after the last day  \\
    & $Vol_{max}$  & Highest recorded DEX daily volume  & $Pval_{max}$  &  Highest recorded LP's total value  \\
    & $Vol_{min}$  & Lowest recorded DEX daily volume   & $Pval_{min}$  & Lowest recorded LP's total value  \\
    & $RVol_{fonl}$  & The ratio of $Vol_{f}$  and $Vol_{l}$ & $RPval_{fonl}$  & The ratio of $Pval_{f}$  and $Pval_{l}$ \\
    & $RVol_{fonmin}$  & The ratio of $Vol_{f}$  and $Vol_{min}$ & $RPval_{fonmin}$  & The ratio of $Pval_{f}$  and $Pval_{min}$\\
    & $RVol_{fonmax}$  & The ratio of $Vol_{f}$  and $Vol_{max}$ & $RPval_{fonmax}$  & The ratio of $Pval_{f}$  and $Pval_{max}$ \\
    & $RVol_{lonmin}$  & The ratio of $Vol_{last}$  and $Vol_{min}$ & $RPval_{lonmin}$  & The ratio of $Pval_{l}$  and $Pval_{min}$ \\
    & $RVol_{lonmax}$  & The ratio of $Vol_{last}$  and $Vol_{max}$ & $RPval_{lonmax}$  & The ratio of $Pval_{l}$  and $Pval_{max}$  \\
    & $RVol_{minmax}$  & The ratio of $Vol_{min}$  and $Vol_{max}$& $RPval_{minmax}$  & The ratio of $Pval_{min}$  and $Pval_{max}$  \\
    \bottomrule
    \end{tabular}
    }
\end{table*}

\end{document}